\documentclass[12pt,preprint]{aastex}
\usepackage{graphicx,verbatim,amsfonts,amsmath,natbib,amssymb}
\usepackage{bm}
\usepackage{times}

\newcommand{\hmsolar}{h^{-1}{\rm\,M_{\sun}}}

\newcommand{\hmpc}{h^{-1}{\rm\,Mpc}}
\newcommand{\hkpc}{h^{-1}{\rm\,kpc}}

\newcommand{\ihmpcC}{h^3 {\rm\,Mpc}^{-3}}
\newcommand{\kmsmpc}{{\rm\ km\ s^{-1}\ Mpc^{-1}}}
\newcommand{\kms}{\ensuremath{\rm\ km\ s^{-1}}}

\newcommand{\om}{\Omega_m}

\newcommand{\ol}{\Omega_\Lambda}

\newcommand{\sqdeg}{{\rm deg}^2}
\newcommand{\zph}{z_{\rm ph}}
\newcommand{\bgc}{B_{\rm gc}}

\newcommand{\eg}{e.g.,}

\newcommand{\etal}{et al.}
\newcommand{\beq}{\begin{equation}}
\newcommand{\eeq}{\end{equation}}
\newcommand{\beqa}{\begin{eqnarray}}
\newcommand{\eeqa}{\end{eqnarray}}

\newcommand{\rvir}{r_{\mathrm 200}}
\newcommand{\bgcunits}{\,(h_{\rm 50}^{-1}{\rm\,Mpc})^{1.77}}

\shorttitle{Color Bimodality in Galaxy Clusters}
\shortauthors{Loh et al.}
\begin{document}
\title{The Color Bimodality in Galaxy Clusters since $z \sim 0.9$ \footnote{
based on observations carried out at the Canada-France-Hawaii Telescope (CFHT), operated 
by the National Research Council of Canada, the Centre National de la Recherche 
Scientifique of France and the University of Hawaii. 
}~$^{,}$\footnote{based on observations taken at the Cerro Tololo Inter-American Observatory}}

\author{Yeong-Shang Loh}
\affil{Department of Physics and Astronomy, University of California, 
  Los Angeles, CA 90095, USA}
\email{yeongloh@astro.ucla.edu}

\author{E. Ellingson}
\affil{Center for Astrophysics and Space Astronomy, University of Colorado, 
  Boulder, CO 80309, USA}

\author{H.K.C. Yee and D.G. Gilbank}
\affil{Department of Astronomy and Astrophysics, University of Toronto, Toronto, ON M5S 3H4, Canada}


\author{M.D. Gladders}
\affil{Department of Astronomy and Astrophysics, University of Chicago, Chicago, IL 60637, USA}

\and

\author{L.F. Barrientos}
\affil{Departamento de Astronom\'{\i}a y Astrof\'{\i}sica,
Universidad Cat\'{o}lica de Chile, Avenida Vicu\~na Mackenna 4860, Casilla 306, Santiago 22, Chile}

\begin{abstract}
We present the evolution of the color-magnitude distribution of galaxy clusters from 
$z = 0.45$ to $z = 0.9$ using a homogeneously selected sample of $\sim 1000$ 
clusters drawn from the Red-Sequence Cluster Survey (RCS). The red fraction of galaxies 
decreases as a function of increasing redshift for all cluster-centric radii, consistent with 
the Butcher-Oemler effect, and suggesting that the cluster blue population may be identified with newly infalling galaxies.
We also find that the red fraction at the core has a 
shallower evolution compared with that at the cluster outskirts. 
Detailed examination of the color distribution of blue galaxies suggests that they have colors
consistent with normal spirals and may redden slightly with time. 
Galaxies of starburst spectral type contribute less than $5$\% of the increase in the blue population 
at high redshift, implying that the observed Butcher-Oemler effect is
not caused by a unobscured starbursts, but is more consistent with a
normal coeval field population.
\end{abstract}

\keywords{methods: statistical -- galaxies: elliptical and lenticular -- galaxies: evolution -- galaxies: clusters: general}

\section{Introduction}\label{sec:intro}
The comparison of local and high redshift clusters of galaxies is a powerful tool for
constraining theories of structure formation and cosmological world models.
The discovery that galaxy clusters at $z \sim 0.5$ have a larger fraction of blue galaxies 
compared with similar clusters found in the local universe \citep{But78,But84} 
provided the first direct evidence of galaxy evolution
in dense environments. Since these early studies, the Butcher-Oemler effect has been
confirmed photometrically \citep{Rak95,Mar00,Mar01,Kod01,Got03} and 
spectroscopically \citep{Dre83,Lav88,Fab91,Dre92,Pog99,Ell01,Pog06}. 
This observed evolution has been interpreted variously as evidence for
a rapid change in galaxy populations driven by ram-pressure stripping \citep{Gun72},
galaxy-merging or other physical processes in clusters \citep{Lav88,Moo96}, as
well as the effect of infalling galaxies from the field or near-field
regions \citep{Kau95,Abr96,Ell01}.

However, the Butcher-Oemler phenomenon has been challenged from the start \citep[e.g.,][]{Koo81}
and more recently by a host of new analyses, drawn primarily from 
X-ray selected samples of clusters and near-infrared observation of 
galaxies \citep{And98,And99,Sma98,Dep03}. In these, the evolutionary trend with redshift was 
found to be either weaker, or non-existent. The 
non-uniformity of the original Butcher-Oemler clusters, especially in physically 
related properties like galaxy surface number density \citep{New88} and X-ray 
luminosity \citep{And99} may present selection biases that lead to a strong 
redshift evolutionary trend. Indeed, the original Butcher-Oemler clusters were 
selected 
from photographic plates and may include lower mass and
unvirialized merging systems. The strong \emph{k}-correction pushes the sample to bluer rest bands. 
These optically-selected galaxy samples may also preferentially select
systems with strong star-formation \citep[e.g.,][]{Gun86,And05}. 

Earlier studies also suffered from low number statistics. While the analyses were usually
done with fewer than $50$ clusters over a large lookback time (e.g., from $z = 0$ to $z = 0.5$),
general claims were often made about the cluster population as a whole. Part of the reason
is that investigators necessarily analysed whatever systems that were available. Uniformly selected 
cluster samples that span a large redshift range and have constant physical attributes 
(e.g., similar masses) are sparse and the variation of blue fraction within a fixed epoch is often 
considerable \citep[e.g.,][]{Mar01}. It is unclear to what extent this variation is due to 
actual cosmic variance or the perculiarities of the sample selection itself, since the number of 
clusters involved is at most a few tens. 

With the advent of wide area photometric and spectroscopic redshift surveys like the
Sloan Digital Sky Survey \citep[][SDSS]{Yor00} and the Two-Degree Field Galaxy Redshift 
Survey \citep[][2dFGRS]{Col01}, large catalogs of uniformly selected clusters are now available at low 
($z \sim 0.1$) to intermediate ($z \sim 0.3$) redshift 
\citep[e.g.,][]{Ann99,Got02,Kim02,Mer02,Eke04,Yan05,Mer05,Wei05,Mil05,Ber06}.
However, in order to understand what processes are dominant in explaining the Butcher-Oemler effect, 
a large redshift range is desirable to provide maximal leverage to evolutionary studies
as various theoretical models may have different redshift dependences.

In this work, we revisit the Butcher-Oemler phenomenon using the currently largest sample of 
uniformly selected clusters of galaxies available, out to $z \approx 0.9$. The galaxy 
clusters are drawn from the complete and well-calibrated Red-Sequence Cluster 
Survey \citep[RCS;][]{Gla05}. The RCS method searches for galaxy overdensities 
along a color-magnitude sequence corresponding to cluster early-type
galaxies, the cluster "red sequence".  The apparent color of this
sequence reddens with redshift, allowing us to separate overlapping
structures and effectively remove most of the  contamination that has
plagued monochromatic cluster surveys.  All massive clusters discovered
by any methodology to date show a detectable red sequence, so that,
while it is slightly less effective for very blue systems,  this
methodology is not expected to introduce a strong bias to the study of
the cluster populations. To the extent that galaxies on the early-type red sequence 
have the largest fraction of 
stellar mass at a fixed luminosities \citep[e.g.,][]{Bel04}, our cluster sample can be considered 
stellar mass dependent. Our sample may exhibit selection differences when compared, 
for example, with dark mass dependent weak lensing detected cluster samples such as the Deep 
Lens Survey \citep[DLS;][]{Wit02}, intracluster medium (ICM) density dependent X-ray 
selected samples \citep[\eg;][]{Ros98}, or star-forming galaxy biased blue overdensities for Lyman break selected 
samples. However, galaxy density maps derived from red sequence galaxies have been shown to 
correlate with weak lensing shear maps with high significance \citep{Wil01}.

In this paper we study the evolutionary trend of the red fraction for cluster galaxies with 
redshift within a range of radii, scaled by $r/\rvir$ -- the radius at which the galaxy 
overdensity is $200$ times the mean density. Scaling the radii to $r/\rvir$ allows 
a statistical combination of many systems and minimizes many
problems arising from the foreground/background subtraction in photometric surveys.
Our approach is cluster-centric \citep[e.g.,][]{Mel77,Whi93,Ell01}, where we use a characteristic
radius $\rvir$ to normalize the clusters across the richness range of the clusters. 
This approach is complementary to many more recent studies of the Color 
(Star Formation Rate)-Density relation \citep{Hog03,Bal04,Yee05} which use local galaxy 
densities \citep{Dre80,Smi05,Pos05,Cuc06,Coo06}. 

The outline of this paper is as follows. In the next section, we
describe the sample used in our analysis. The procedure for the statistical 
background correction of composite clusters is described in section~\ref{sec:methods}.  
In section~\ref{sec:discussion}, we discuss our results  
and give caveats that need to be taken into account for robust interpretation. 
In section~\ref{sec:summary}, we summarize our conclusions. 
Throughout this paper, $h=H_0/100\kmsmpc$ and we adopt the 
concordance $\om = 0.3, \ol = 0.7 $ cosmology.
\section{Sample}\label{sec:sample}
\subsection{The RCS Survey}
The Red-Sequence Cluster Survey \citep[RCS;][]{Gla05} is a $\sim 100\,\sqdeg$ imaging survey in the $z'$ and $R_c$ 
bandpasses conducted with the CFHT 3.6 m and CTIO 4 m with the primary goal 
of searching for clusters of galaxies up to $z \approx 1.4$. There are $22$ RCS patches 
(each of $\approx 2.5\, \sqdeg$) distributed over the northern and southern sky.
Some of the patches were chosen to overlap with other widely studied fields, such as 
XMM-LSS \citep{Pie04}, CDF-S \citep{Gia01}, CNOC2 \citep{Yee00}, SWIRE \citep{Lon03} and PDCS \citep{Pos96}. 
\subsection{Photometry, Calibration, Extinction}
The primary RCS photometric reduction of object detection, source deblending, 
star-galaxy separation and aperture photometry was performed using the Picture 
Processing Package \citep[PPP;][]{Yee91,Yee96}. The detailed implementation of the RCS 
photometric pipeline is described in \cite{Gla05}. The average 
$5 \sigma$ limiting magnitudes for point sources are $z'= 23.9$ and 
$R_c = 25.0$ (AB system). Analysis of log(N)-log(S) galaxy counts suggests that the 
photometry for extended sources is complete to $\sim 0.8$ mag brighter, although there are variations from pointing 
to pointing. There are over $3$ million point sources and $11$ million galaxies in the RCS database.
The relative photometry between patches is better than $0.1$ magnitudes \citep{Hsi05}. 
Each detected extended source is further corrected for Galactic extinction, interpolating the
dust map of \cite{Sch98} and assuming $R_{R_c} = 2.634$ and $R_{z'} = 1.479$. 
For the present analysis, we use $72\,\sqdeg$ of the
RCS survey with the best photometry \citep{Gla06}. Throughout this paper, our $R_c$ and $z'$
photometry is presented on the AB system.
\subsection{The Cluster Sample}
About $8000$ groups and clusters of galaxies were detected with high significance by searching 
for source overdensities in color slices along the red-sequence of early-type galaxies. Details 
of the cluster detection method are outlined in \cite{Gla00}. The release of the first catalog, 
as well as improvements over the cluster detection methodology, are described in \cite{Gla05}. 
A useful by-product of the red-sequence method is an estimated photometric redshift for each 
detected cluster. RCS uses only two bandpasses strategically chosen to straddle 
the $4000$\AA~break of the target redshift coverage, and photometric redshifts of the detected 
clusters are quite well constrained, to $\Delta z/ z \approx 0.1$ \citep{Gil07s}. 
The dense collection of red galaxies along the red-sequence effectively diminishes the 
Poisson noise from photometric uncertainties of individual sources (\citealp{Lop04}).

In addition to an estimated photometric redshift (hereafter $\zph$), an optical 
richness parametrized by the amplitude of the galaxy-cluster cross-correlation function 
$\bgc$ \citep{Lon79,Yee87,Yee99} is measured. $\bgc$ is found to correlate well with cluster 
properties such as X-ray temperature and galaxy velocity dispersion at intermediate redshift, and 
it is relatively well calibrated for X-ray selected clusters at $z \sim 0.35$ \citep{Yee03}. 
Note that here we use red sequence galaxies for the measurement of $B_{\rm gc}$ \citep{Gla05},
which gives a tighter relation with respect to X-ray properties, especially at higher 
redshift \citep{Hic07}. ROSAT observations have also found that most groups and lower mass systems with 
extended X-ray halos are dominated by early-type galaxies \citep{Osm04,Mul03}. $B_{\rm gc}$
computed from the red sequence will down-weight unvirialized poor systems without a hot ICM. 

\begin{figure*}
\includegraphics[width=0.48\textwidth,height=0.45\textwidth]{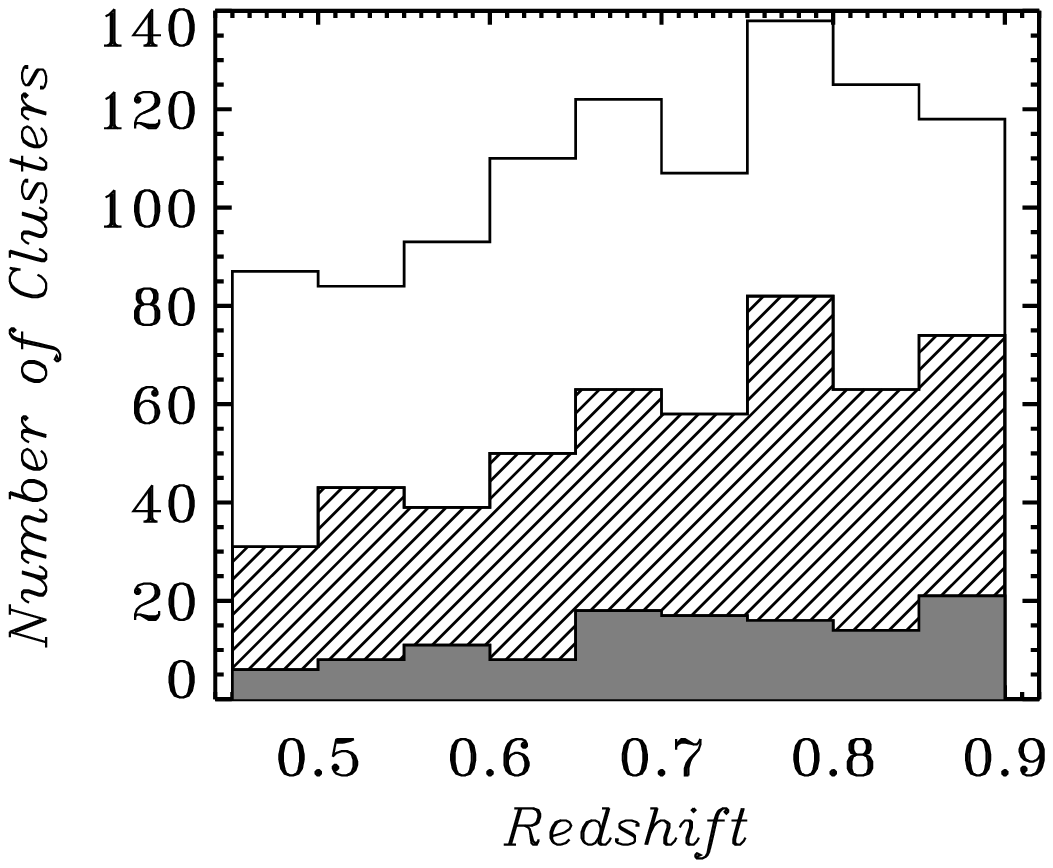}
\caption{
The redshift histogram of RCS clusters used in our analysis. The three segments indicate richness bins of
$\bgc > 800$ (grey for rich clusters), $500 < \bgc < 800$ (hatched) and $300 < \bgc < 500$ (white for poor clusters). 
\label{fig:hist}}
\end{figure*}

For the present analysis, we use $\sim 1000$ clusters with $B_{\rm gc} > 300 \bgcunits$ and 
$0.45 < z_{\rm ph} < 0.90$. Mass calibration using low to intermediate redshift clusters 
suggests these systems have richnesses corresponding to Abell Richness Class (ARC) $\gtrsim 0$, 
$\sigma_{v} > 450 \kms$ and $T_X > 3~{\rm keV}$ \citep{Yee99,Yee03}. The sample is divided into 
five redshift slices with $171$ to $263$ systems in each slice (Table \ref{tab:sample}). 
While the mass of each cluster is uncertain due to the rather large uncertainty in the measured 
$B_{\rm gc}$ and the intrinsic scatter in mass-$B_{\rm gc}$ relation,
the overall distribution of cluster counts as a function of redshift is robust. 
Figure \ref{fig:hist} shows the redshift histogram of clusters used in our analysis. 
Analysis of the RCS cluster counts to constrain cosmological 
parameters yields results consistent with the concordance values \citep{Gla06}, indicating that
large systematic errors in our mass estimates are unlikely. 
From abundance matching, clusters with $B_{\rm gc} > 300 \bgcunits$ used in this paper are rare, 
with comoving densities $< 10^{-5} \ihmpcC$, consistent with dark matter halos with mass 
$\gtrsim 10^{14} \hmsolar$.
\section{Methodology} \label{sec:methods}
\subsection{Foreground and Background Subtraction}
\begin{figure*}  
\includegraphics[width=0.32\textwidth,height=0.32\textwidth]{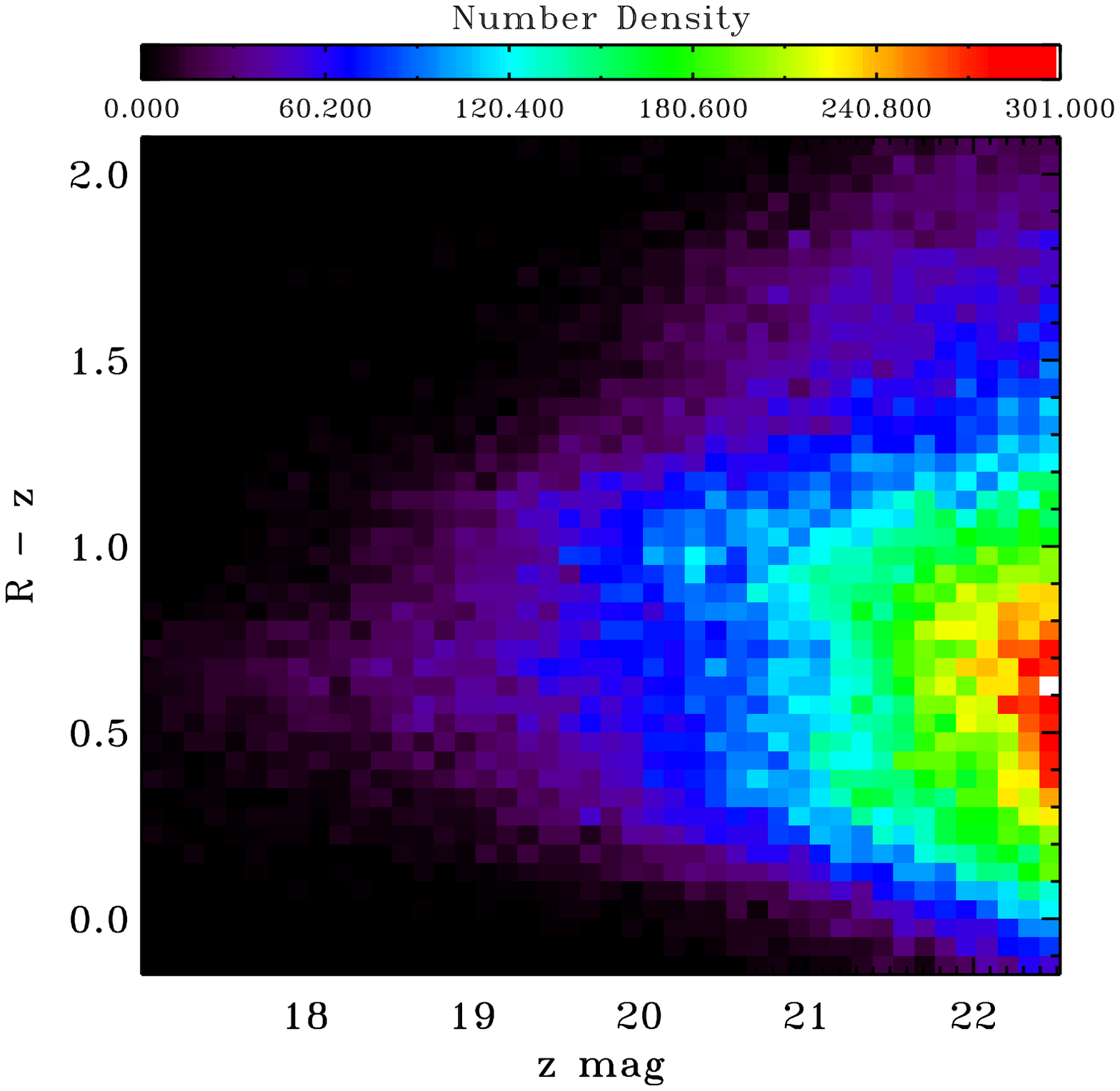}
\includegraphics[width=0.32\textwidth,height=0.32\textwidth]{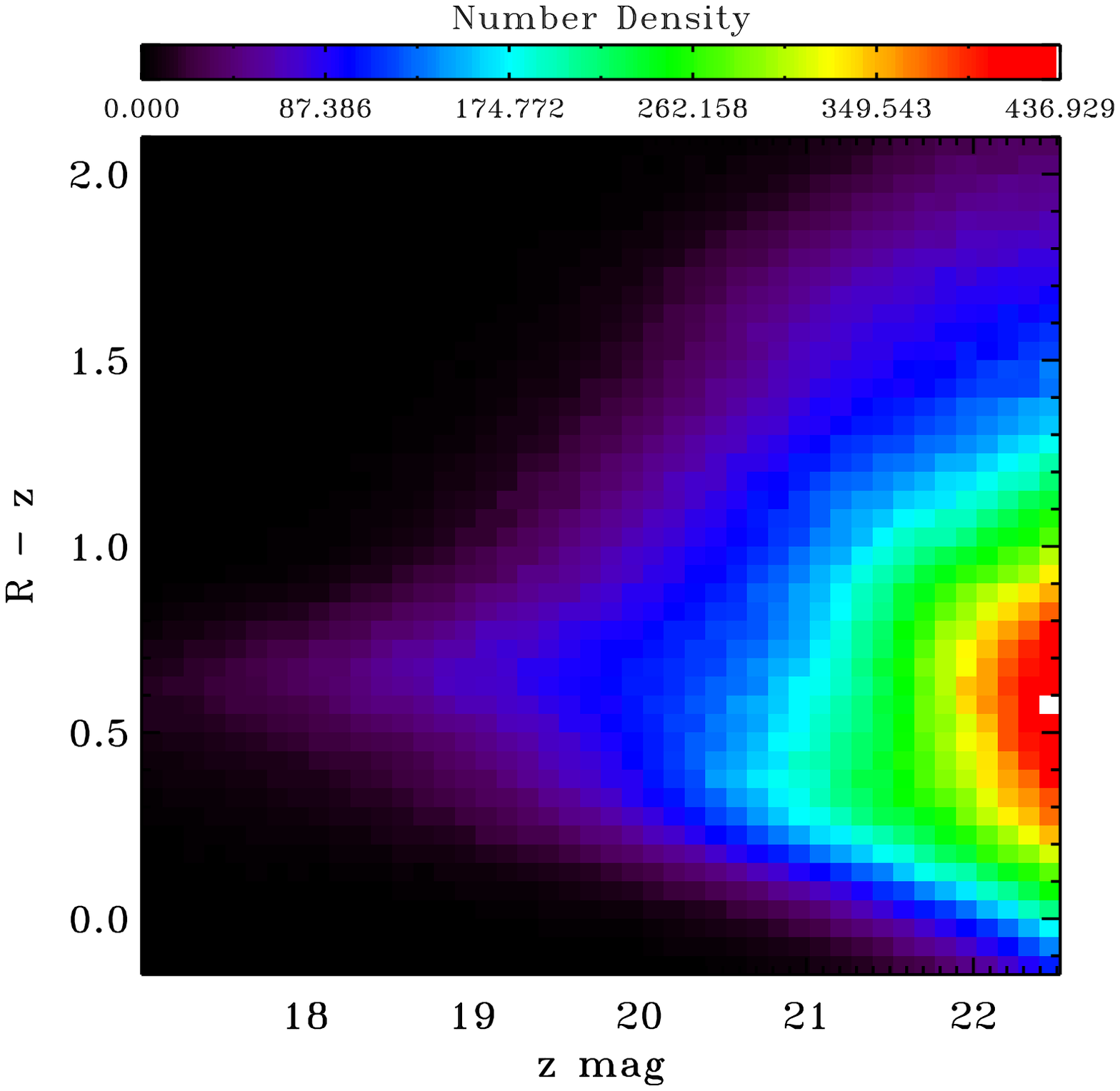}
\includegraphics[width=0.32\textwidth,height=0.32\textwidth]{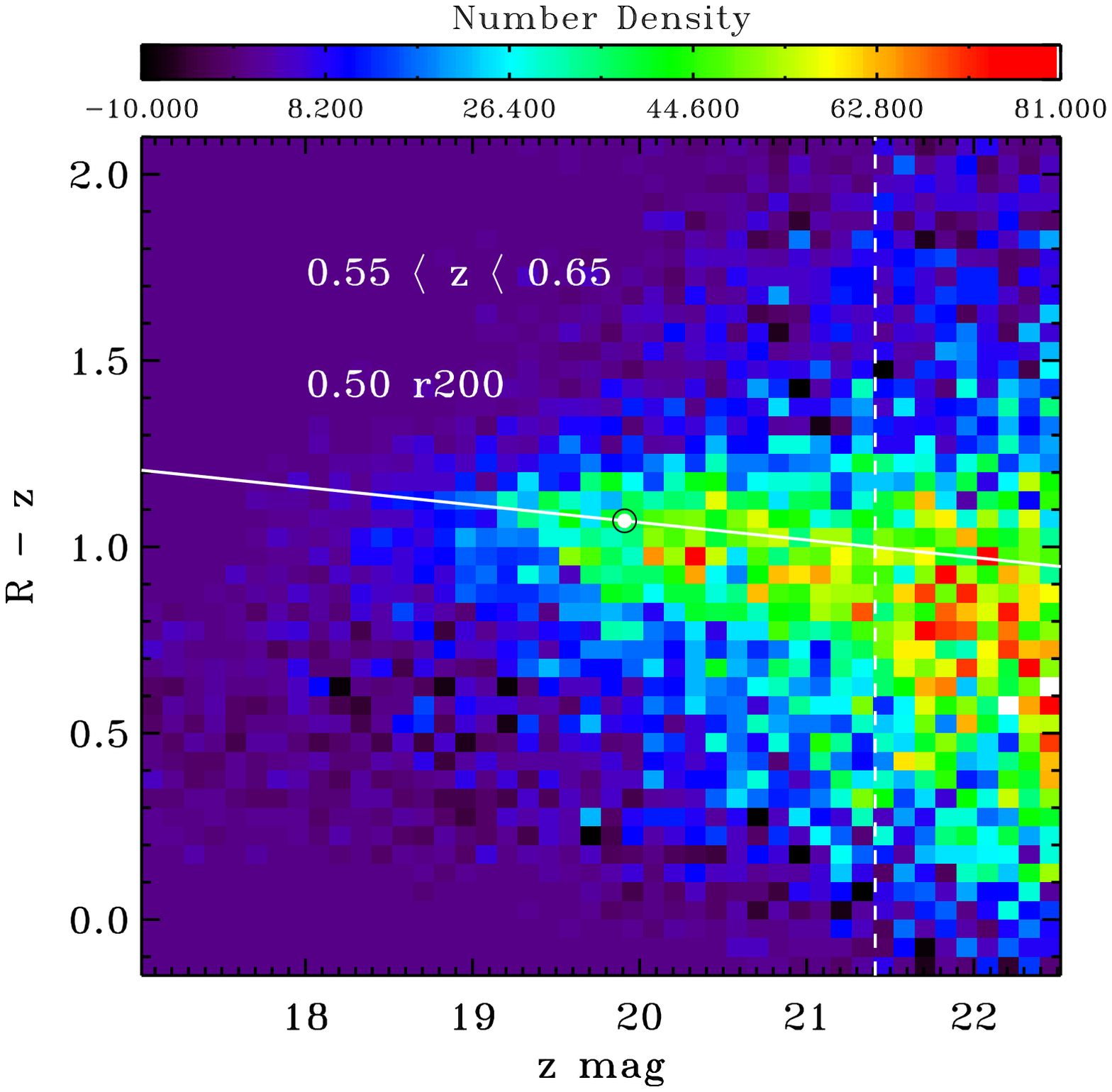}
\caption{
(Left) The observed color-magnitude distribution for all non-stellar sources
with $r < 0.5~\rvir$ for $203$ clusters with $0.55 < \zph < 0.65$. 
(Middle) Patch-weighted background distribution (see text). 
(Right) Background-subtracted distribution of galaxies, calculated from the two preceding panels.
The white dot indicates $M^\star$ at the median redshift ($\sim 0.61$) while the
dashed vertical line indicates $M^\star + 1.5$, the consistent flux limit used for the red fraction
analysis. The completeness of the photometry for extended sources ($z' < 23.1$ and $R_c < 24.2$) 
lies beyond the boundary of these figures. 
\label{fig:color_mag}}
\end{figure*}

To obtain the composite color-magnitude distribution of our cluster galaxies, we have to subtract 
the contribution from foreground and background galaxies statistically. The two-dimensional
observed background counts for non-stellar sources in $R_c - z'$ color and $z'$ 
magnitude are constructed for each of the $22$ RCS patches. Depending on the relative number of 
clusters from each patch used for the composite cluster analysis at each redshift slice, a weighted 
background based on the effective area contributed by the clusters
is created (Figure \ref{fig:color_mag}). This semi-global 
background approach allows us to model the background more accurately using a large area, but still
mimics some inherent patch to patch variations such as observing depth and seeing. This
approach was similarly employed by \cite{Han05} but performed in two dimensions 
(in color and magnitude space) as in \cite{Loh05}. Our approach is in contrast with 
\cite{Val01} who advocate a local background subtraction methodology. \cite{Got03} find 
little difference with the two approaches if a large enough area is used. The global approach 
has the advantage that it uses all available data to constrain a smooth background with high 
precision. For a given patch of survey area, the region occupied by the clusters, after 
aggregating in redshift, takes up $2-5\%$ of the entire patch area. We rescaled the 
background to match this total cluster area before the background subtraction. 

We are interested in the average properties of clusters at a given redshift. If we assume that 
clusters are self-similar systems with a universal galaxy density profile, then clusters of 
different physical sizes can be scaled to some normalized cluster radius. However, without 
X-ray information or spectroscopy, it is difficult to measure the virial radius of each of the 
clusters. We use the measured $B_{\rm gc}$ for each cluster to estimate a virial radius 
$\rvir$ using the relation $\log \rvir = 0.48 \log \bgc - 1.10$. This relation is empirically calibrated 
from intermediate redshift clusters \citep{Yee03,Bli06}.  We stack 
all clusters in a given redshift slice, but scale each cluster by its $\rvir$ for an effective 
normalizing scale. We emphasize here that $B_{\rm gc}$ is measured at the core, within 
$0.25~\hmpc$, and that the self-similar assumption is implicit in \cite{Yee03}. 
We note here that the spherical halo assumption inherent 
in the $B_{\rm gc}$ measurements is not perfect, as in addition to spherical distribution of 
red galaxies, the RCS method may systematically include physically associated filamentary galaxy
structure along the line-of-sight. 
However, because of the overdensities implied at $B_{\rm gc} > 300 \bgcunits$, most structures 
will be dynamically collapsed, if not completely virialized. 
We are exploring secondary measurements 
such as the concentration of the cluster galaxy distribution or the dominance of the BCG as a 
possible second parameter to further refine the mass measurements. 

Figure \ref{fig:color_mag} shows the color-magnitude distributions of 
galaxies with $r/\rvir < 0.5$ at $z \approx 0.6$. We have stacked $203$ clusters with 
$0.55 < \zph < 0.65$ to obtain the observed distribution (left panel). 
The red sequence is clearly visible.
The patch-weighted background distribution is shown next in the middle panel. 
The distribution after background subtraction is shown in the right panel. 
A prominent red sequence appears after the subtraction. 
The solid line is the fitted red sequence 
at the effective redshift. The solid dots indicate the model $M^\star$ at this redshift 
and the dashed vertical line indicates $1.5$ magnitude fainter than $M^\star$. 
\subsection{Semi-Empirical $k+e$ corrections}
\subsubsection{Red Sequence $k+e$ models}
When computing statistics like red (or blue) fractions across a range of redshifts, the use of an
evolving luminosity limit is crucial since the absolute luminosity of galaxies  
changes by about one magnitude from $z = 0$ to $z \sim 1$.
The original \cite{But84} prescription of using a fixed rest-frame $M_V = -20$ 
does not produce a consistent galaxy population across a range of redshifts.
This bias is especially severe for studies which span a large redshift range. The known correlation between luminosity and 
color can also mimic the evolution of the blue fraction. However, there is considerable uncertainty 
in the measurement of luminosity evolution beyond $z = 0.6$. Work by \cite{Fab05} and \cite{Bel04} 
showed that DEEP2 and COMBO-17 surveys yielded very similar results out to $z \sim 1$ with respect to the $B$-band 
luminosity density for red galaxies. However, \cite{Bro07} and \cite{Sca07} found a milder luminosity density evolution in
the NOAO Deep Wide-Field and COSMOS surveys. 

The luminosity function was determined empirically from background-subtracted galaxy
counts in the red sequence of the clusters. A more detailed study of
the RCS cluster luminosity function can be found in \cite{Gil07};
our assumed luminosity function is within $0.1$ {\tt mag} of the values listed in
their Table I.  These values of $M^\star$ (in the $z'$ bandpass) as a function of redshift are
consistent with the combination of $k$-correction for an early-type
galaxy spectrum along with passive evolution from a $z \sim 3$ single
starburst population (see \citeauthor{Gil07} for details).

While our exact value of $M^\star$ may differ
slightly from that measured by others, the evolution in $M^\star$ within our own
sample is robust and will not influence trends with redshift. 
We explored the possible effects of uncertainty in the cluster galaxy
luminosity function.  A systematic change of $\pm 0.25$ {\tt mag} per redshift in $M^\star$ for the
cluster galaxies produced a change in red fraction less than $1\sigma$.

Our goal is to obtain a consistent rest-frame flux for the cluster galaxies across the redshift
range we are probing. We first use our model $M^\star$ to convert the apparent $z'$ magnitudes of 
red sequence galaxies to evolution-corrected luminosities at the redshift of the cluster.     
The existence of the red sequence then allows us to match the color-luminosity distribution 
across redshift. The observed slope can be used to change the apparent $R - z'$ color into units of 
$C^\star$ ($R - z'$ color at $M^\star$), without appealing to model dependent color evolution 
\citep[e.g.,][]{Bel04,Loh05}. Figure \ref{fig:color_mag2} shows the 
color-luminosity distribution for two redshift slices: $0.45 < \zph < 0.55$ 
(first panel from the left) and $0.85 < \zph < 0.90$ (3rd panel). 
The axes are in units of $M^\star$ and $C^\star$, the color relative to the red sequence.
\subsubsection{Differential $k$-correction}\label{sec:diffk}
While model independent, the approach from the above section 
does not take into account the differential $k$-corrections 
between galaxies of different spectral types for a fixed observed $z'$ magnitude. 
To this end, we use the observed 
color blue-ward of the red-sequence to infer the relative $k$-correction beyond the flux 
relative to a $M^{\star}$ at that epoch, which gives a comparable $k+e$-correction to more 
traditional approaches \citep[\eg;][]{Yee05}

\begin{figure*}
\includegraphics[width=0.24\textwidth,height=0.24\textwidth]{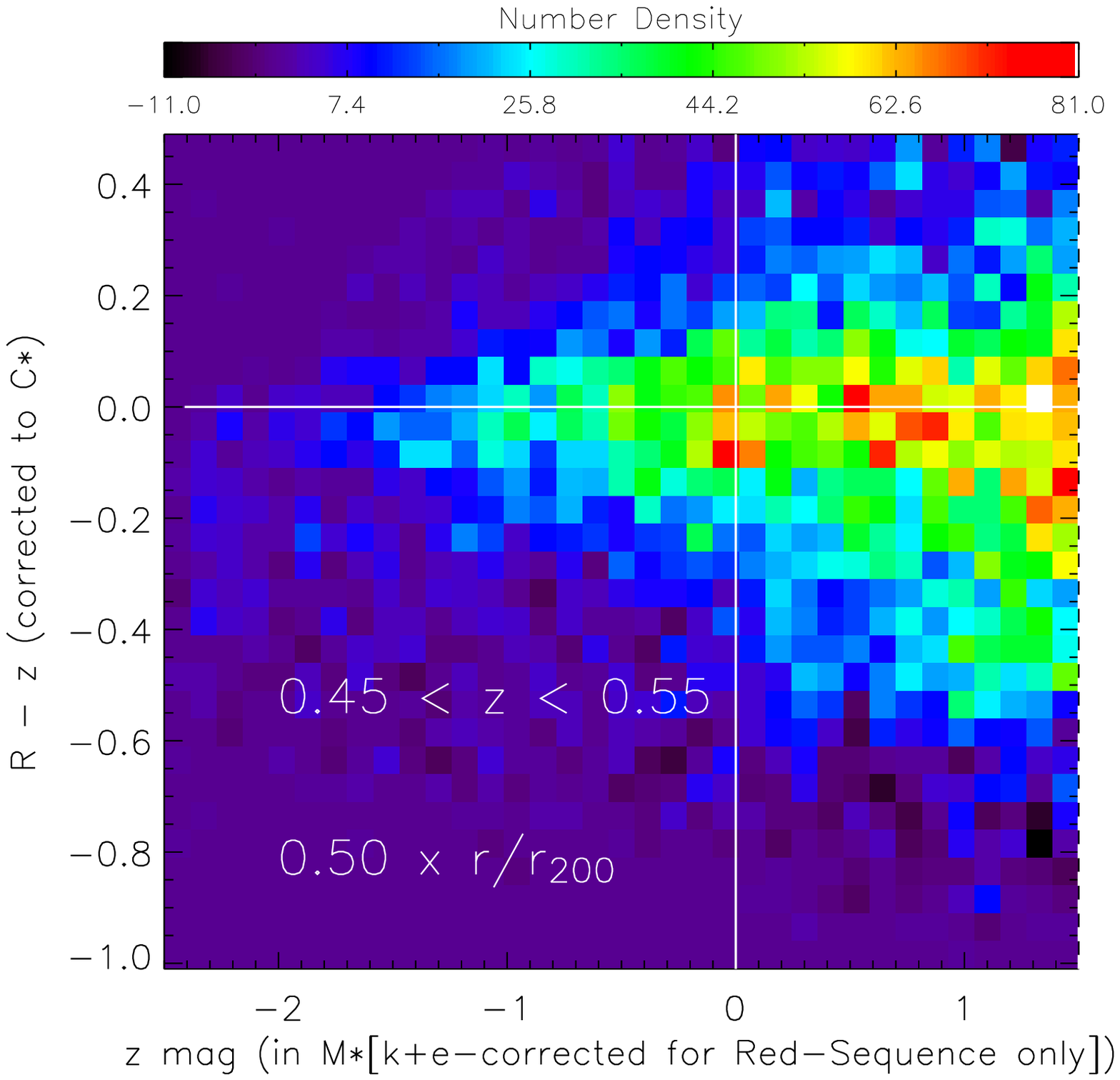}
\includegraphics[width=0.24\textwidth,height=0.24\textwidth]{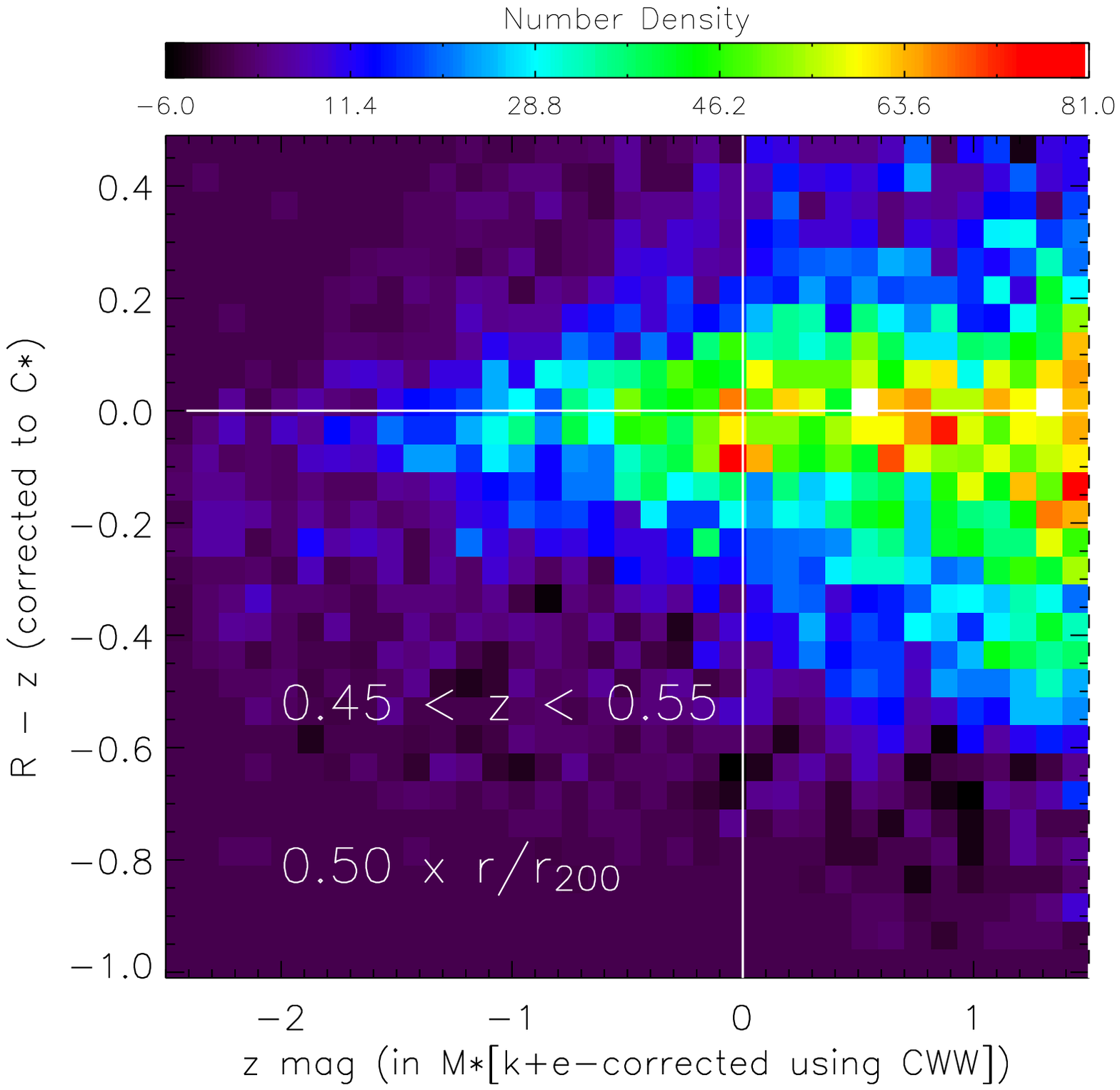}
\includegraphics[width=0.24\textwidth,height=0.24\textwidth]{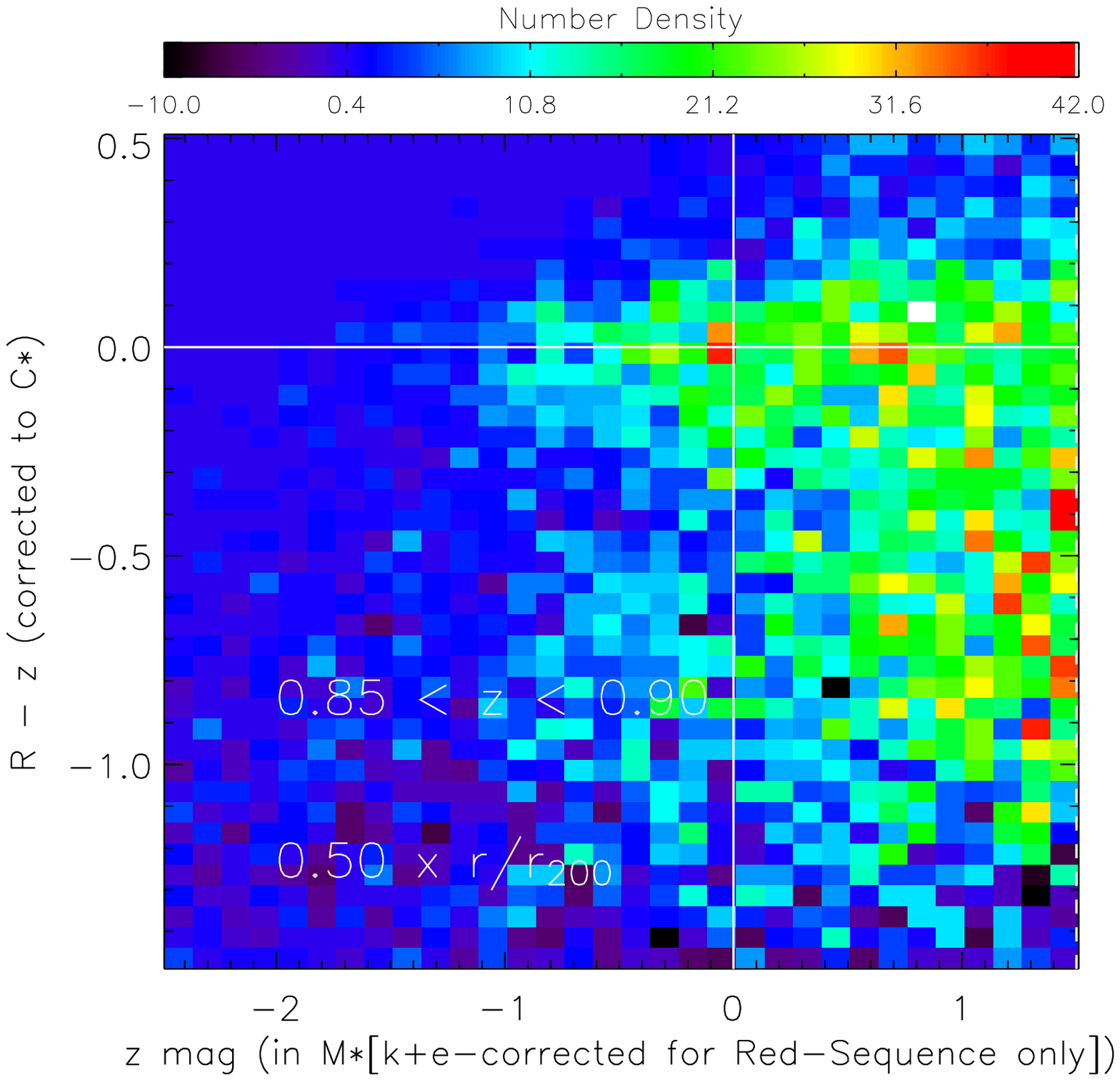}
\includegraphics[width=0.24\textwidth,height=0.24\textwidth]{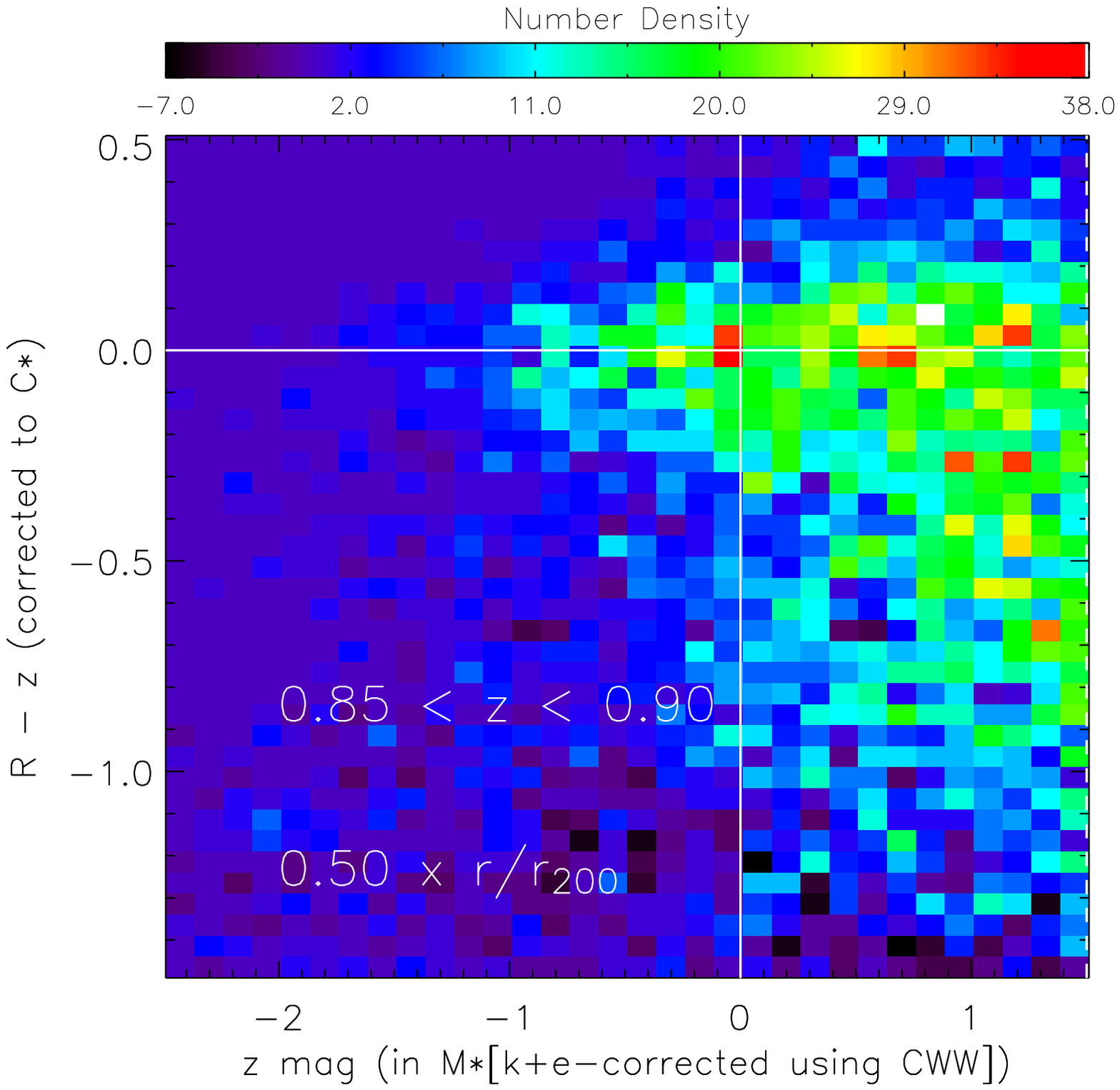}
\caption[]{\label{fig:color_mag2}
Color-magnitude distribution after $k+e$ correction for $0.45 < \zph < 0.55$ (1st and 2nd panel
from the left) and $0.85 < \zph < 0.90$ (3rd and 4th). The axes of the plots are in units of 
$M^\star$ and $C^\star$ at their respective redshifts.
In the 1st and 3rd panel, the $z'$-band flux is corrected consistently for red sequence 
galaxies only. In the 2nd and 4th panel, a differential $k$-correction is based on the expected
color of galaxies of different spectral type. This latter correction not only tightens the red sequence
at low redshift, but also substantially reduces the counts of blue $M < M^\star$ galaxies at high redshift.
}
\end{figure*}

\begin{figure*}
\begin{center}
\includegraphics[width=0.33\textwidth,height=0.29\textwidth]{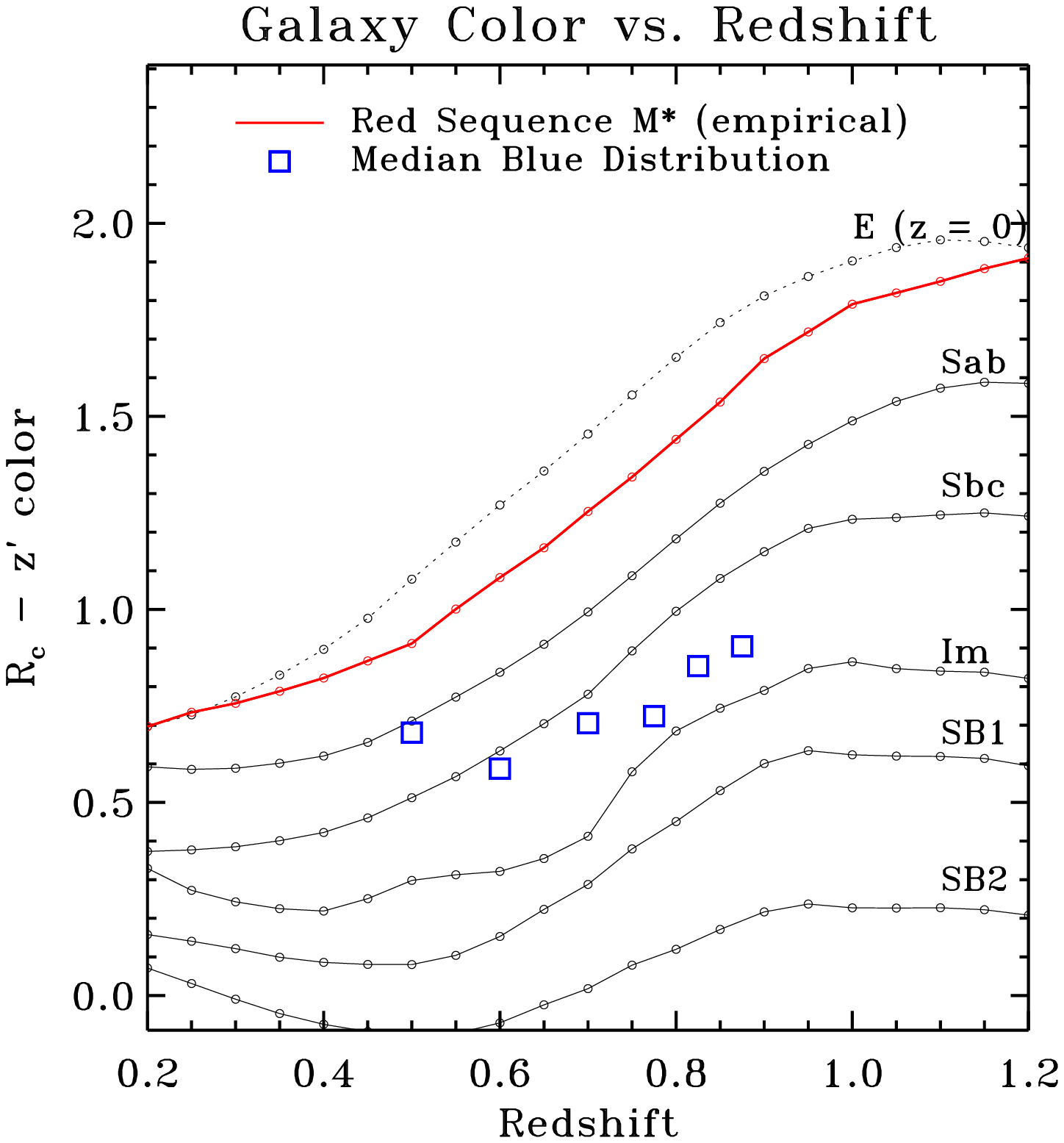}
\includegraphics[width=0.33\textwidth,height=0.29\textwidth]{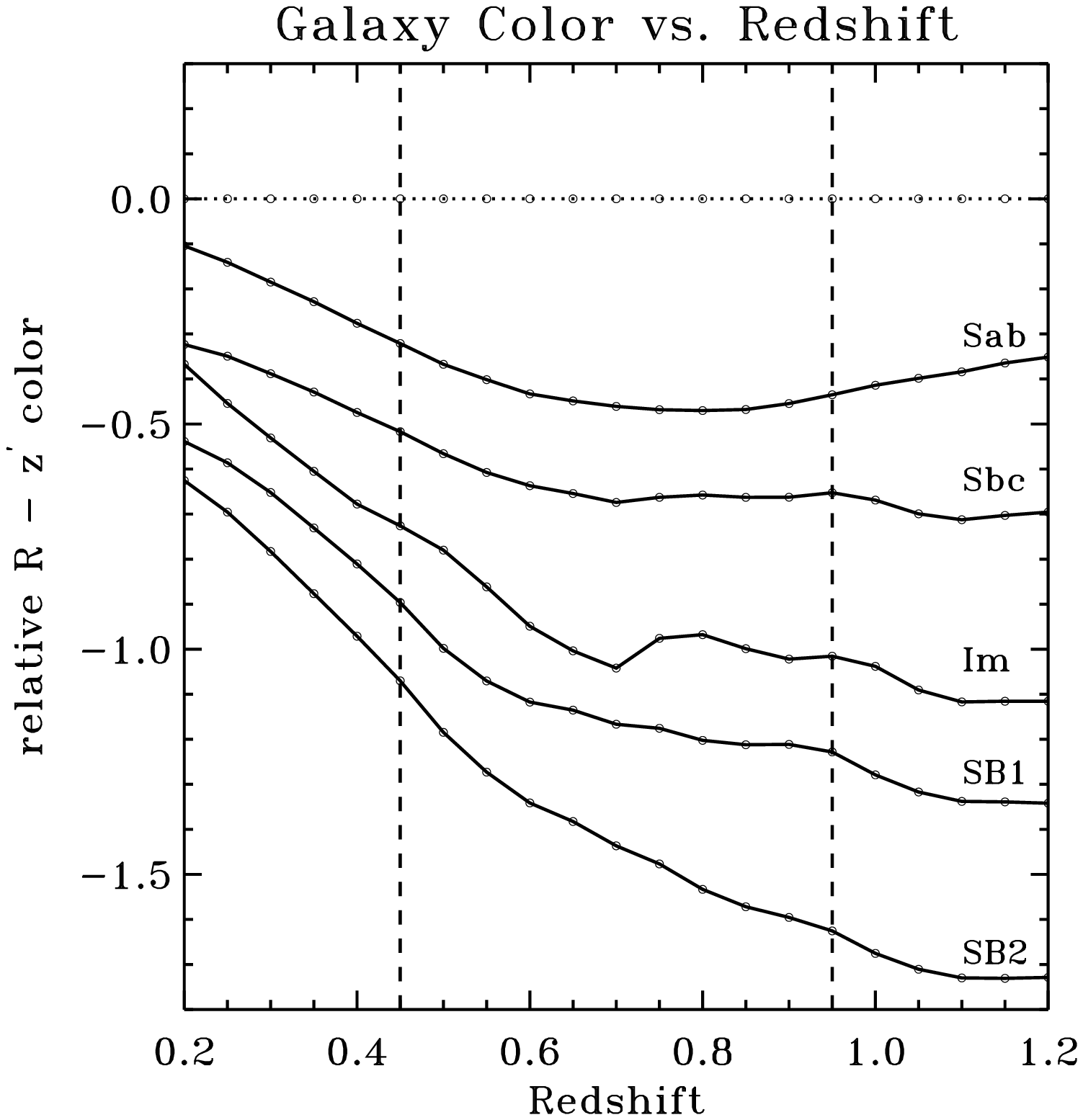}
\caption[]{\label{fig:kcor}
(Left) Synthetic $R_c - z'$ colors for galaxies of different spectral type computed 
using galaxy template spectra from \cite{Col80} and model starburst (SB1, SB2) galaxy spectra. 
The evolving red-sequence $M^\star$ model is shown in red. 
The series of squares are the median color of the non-red-sequence galaxies 
from RCS clusters (see section \ref{sec:blue_gal} and Figure \ref{fig:bdist}) as a function of 
cluster redshift.
(Right) Synthetic colors relative to a fiducial $\sim M^\star$ elliptical galaxy
as a function of redshift. The y-axis is in units of $C^\star$, the color relative to the
colors of ellipticals from the CWW models. The vertical dashed lines bracket the redshift range of our sample. 
}
\end{center}
\end{figure*}

The left panel of Figure \ref{fig:kcor} shows the synthetic $R_c - z'$ color for galaxies of 
different spectral types computed using {\em real} galaxy template spectra from
\cite{Col80} augmented with two starburst models. 
The empirically determined red-sequence color as a function of redshift is shown in red.
Since the red sequence includes passive evolution of the stellar population, it does not match
the non-evolving colors derived from local ellipticals at high redshift.
Plotted on the right panel of Figure \ref{fig:kcor} are the synthetic colors, relative to the 
local ellipticals, for galaxies of different spectral type as a function of redshift. 
The y-axis can be compared directly with the y-axis of the red-sequence corrected color-magnitude 
distribution of Figure \ref{fig:color_mag2}.
From the relative color of each galaxy, we assign a spectral type to each, 
interpolating if the color is in between types. The relative $k$-correction in $z'$-band, 
used to shift a given galaxy to a lower intrinsic luminosity, is inferred from the differential 
$k$-correction relative to the fiducial red sequence template at each reshift. 

Figure \ref{fig:color_mag2} shows the changes in the color-magnitude 
distributions before (1st and 3rd panels) and after (2nd and 4th panels) applying the relative 
$k$-correction to the $z'$-band flux. This additional relative $k$-correction corrects 
blue galaxies for on-going star-formation by pushing them to fainter magnitudes, such that the 
bluest faint galaxies, which would otherwise just make our luminosity cut, are pushed below the 
magnitude limit and are so removed from our sample.
This effect is clearly visible at higher redshift by examining blue
galaxies in the 3rd and 4th panels of Figure \ref{fig:color_mag2} ($0.85 < \zph < 0.90$).  
The numerous blue galaxies brighter than $M^\star$ are corrected to fainter
magnitudes but are still bright enough to be included in the sample. If we
neglect the differential evolution between red sequence and blue galaxies, the derived red fraction
at $\rvir$ is reduced by $2$\% at $z \sim 0.5$ and by $14$\% at $z \sim 0.8$. 


\subsection{Cluster Red Fractions}
There are a number of different ways in which red or blue galaxy fractions are defined in the literature. 
Some use a fixed rest-frame color cut, say rest-frame $B-V < 0.8$ {\tt mag} for blue galaxies 
\citep[\eg][]{Rak95}. 
Definitions like these usually yield a large number density evolution of early-type galaxies that 
are grossly inconsistent with the hiearchical model of structure formation \citep[\eg][]{Wol03}. 
Others, such as the original \cite{But84} studies, use a fixed color width $\Delta$ from the 
observed red sequence, say, $\Delta (B-V) = 0.2$ {\tt mag} for red galaxies \citep[\eg][]{Bel04}. As 
reviewed by \cite{And05}, a constant width bracketing the red sequence is not 
optimal because galaxies of different spectral types drift in and out of the fixed region $\Delta$ 
for even the most generic models of stellar population. Claims of
the evolution the blue (or red) fraction may just be a consequence of systematic spectral types 
drifting in and out of the width at different redshifts. 

\begin{figure*}  
\includegraphics[width=0.32\textwidth,height=0.32\textwidth]{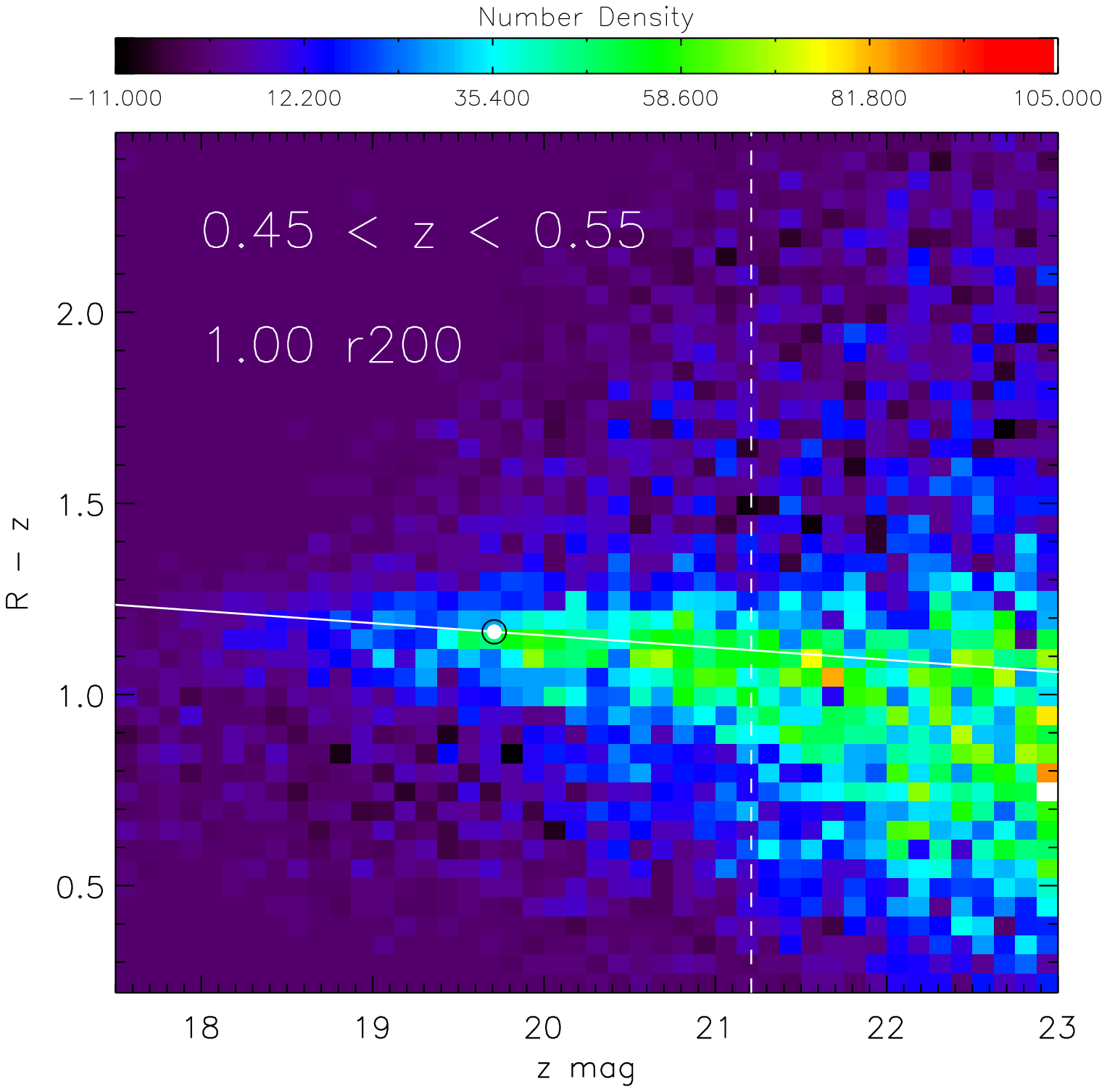}
\includegraphics[width=0.32\textwidth,height=0.32\textwidth]{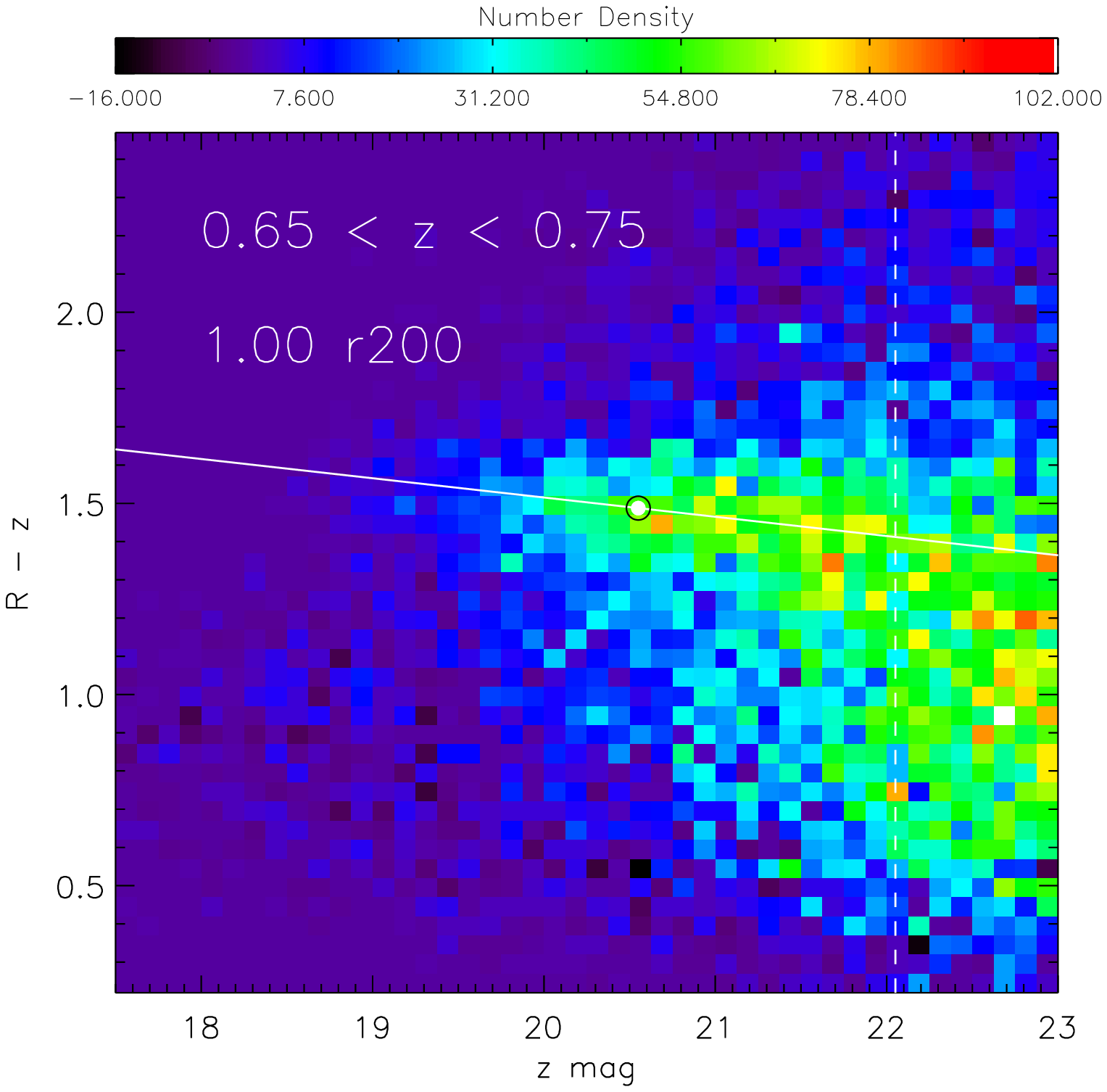}
\includegraphics[width=0.32\textwidth,height=0.32\textwidth]{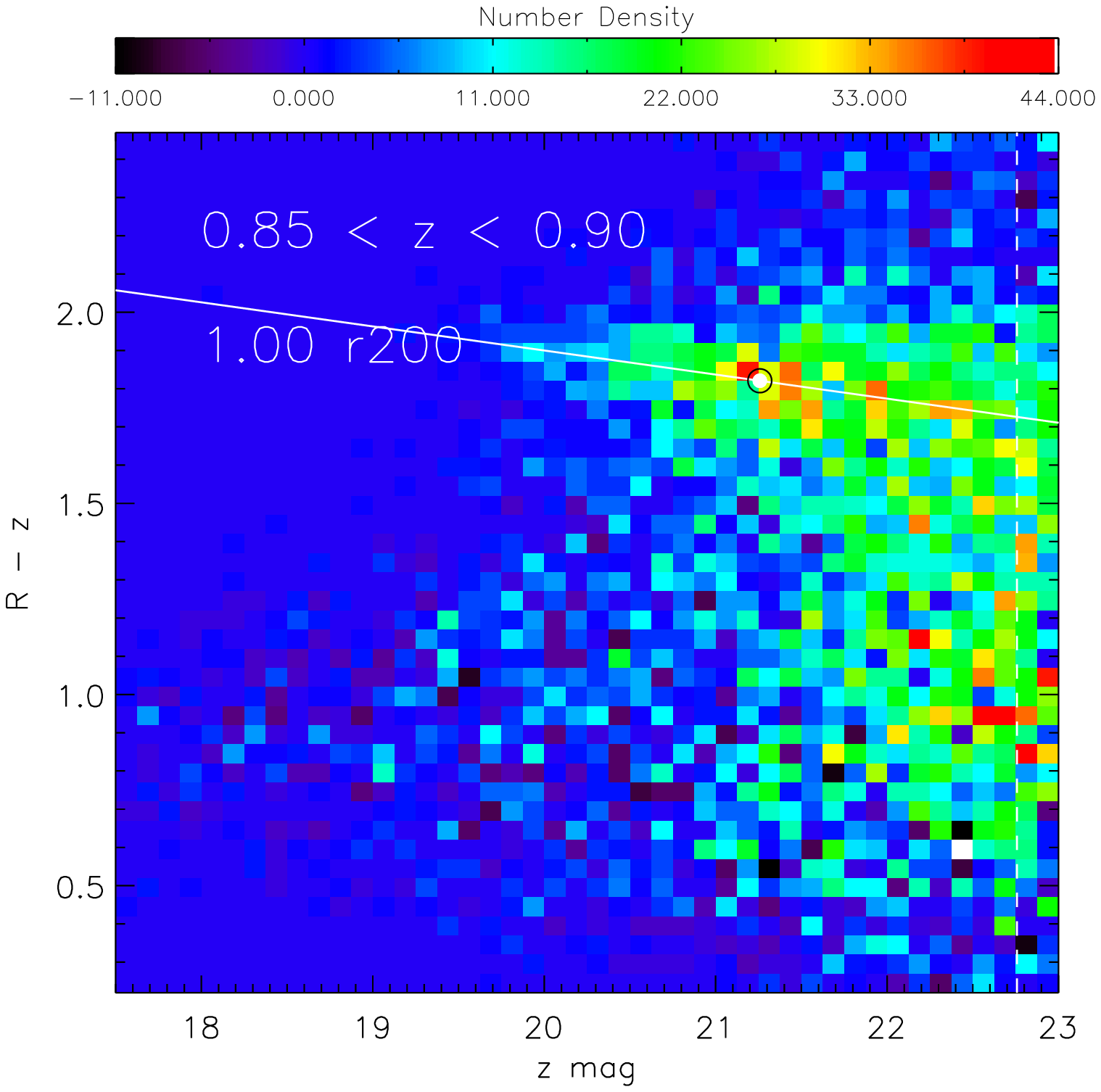}
\includegraphics[width=0.32\textwidth,height=0.32\textwidth,angle=90]{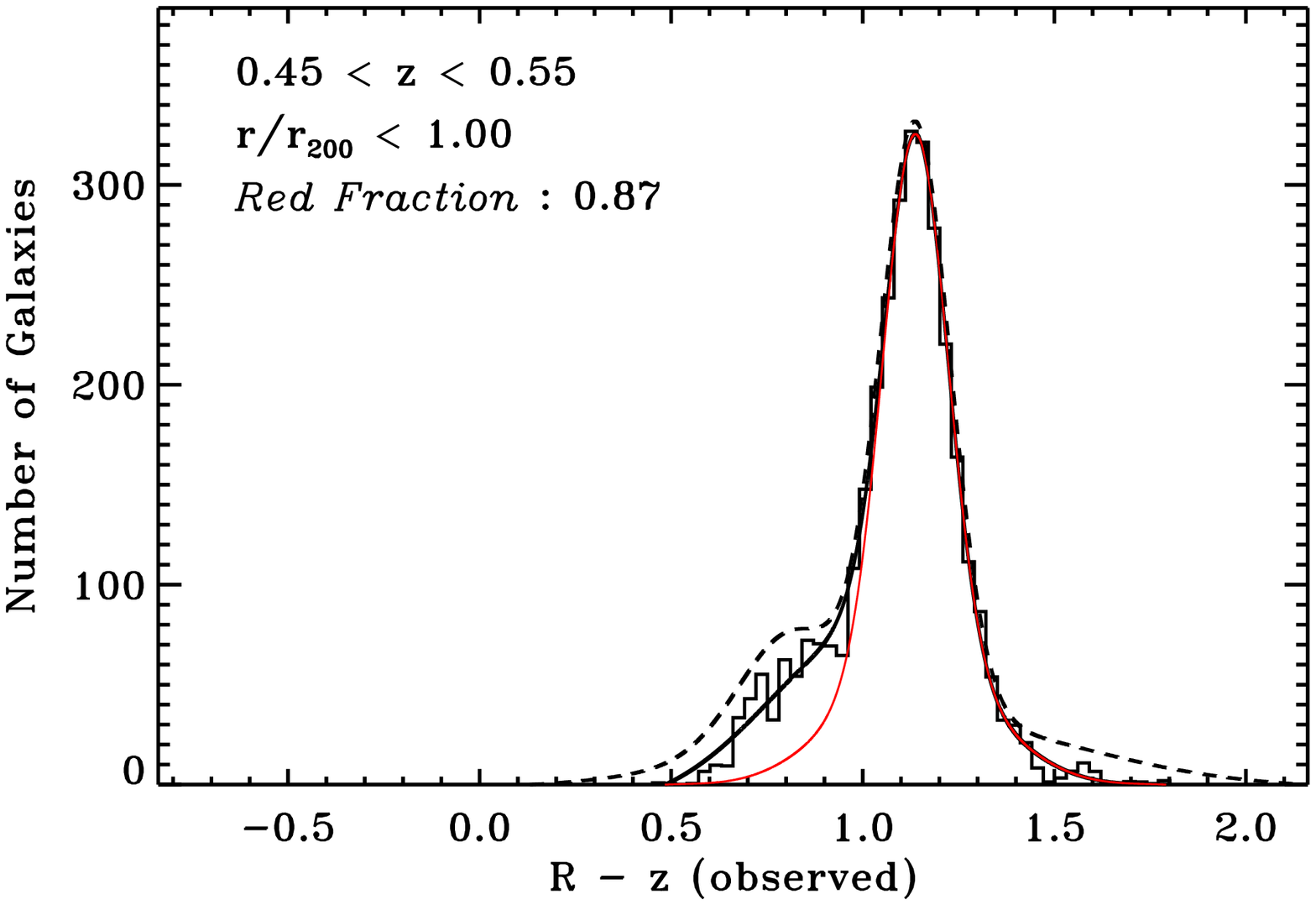}
\includegraphics[width=0.32\textwidth,height=0.32\textwidth,angle=90]{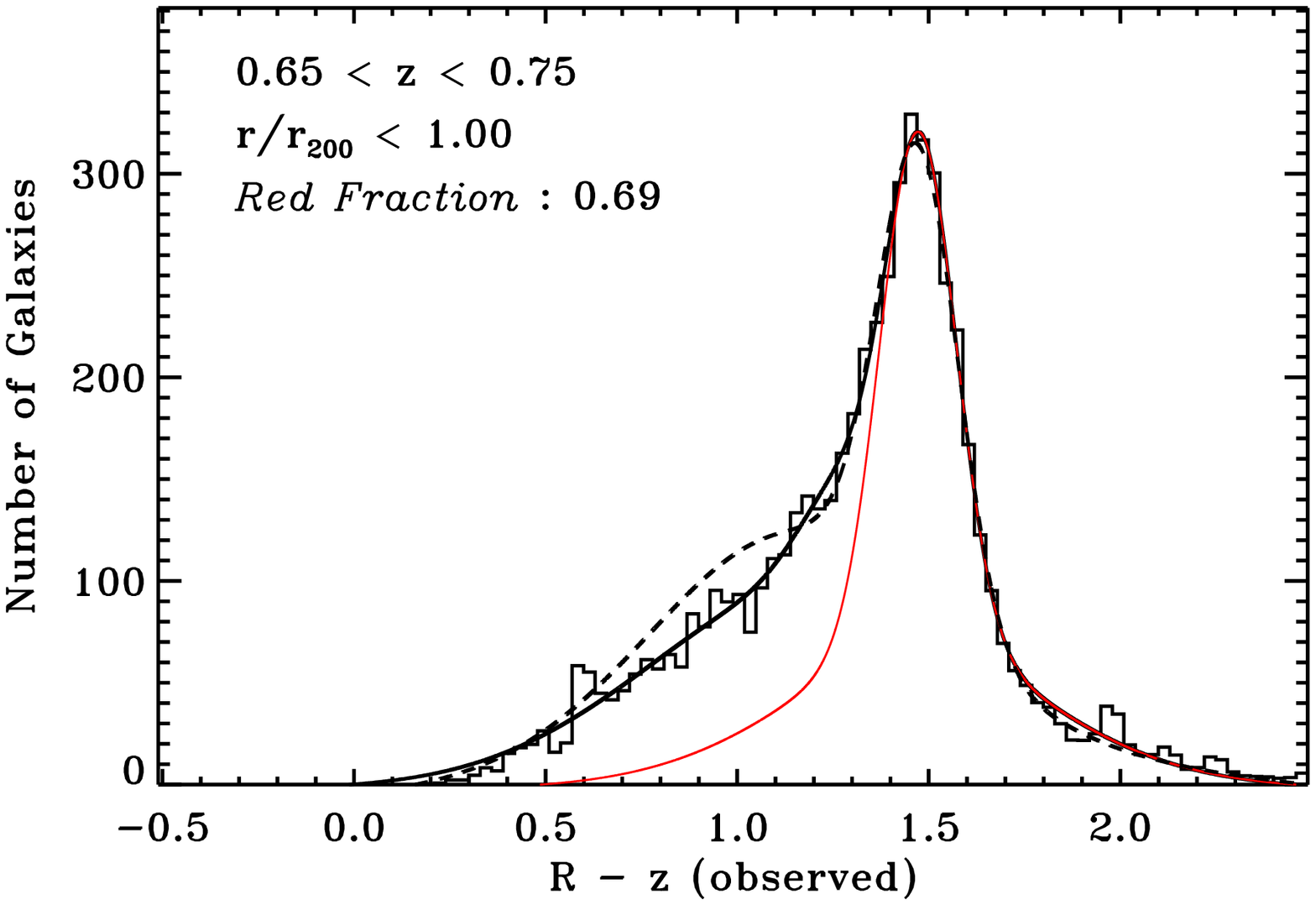}
\includegraphics[width=0.32\textwidth,height=0.32\textwidth,angle=90]{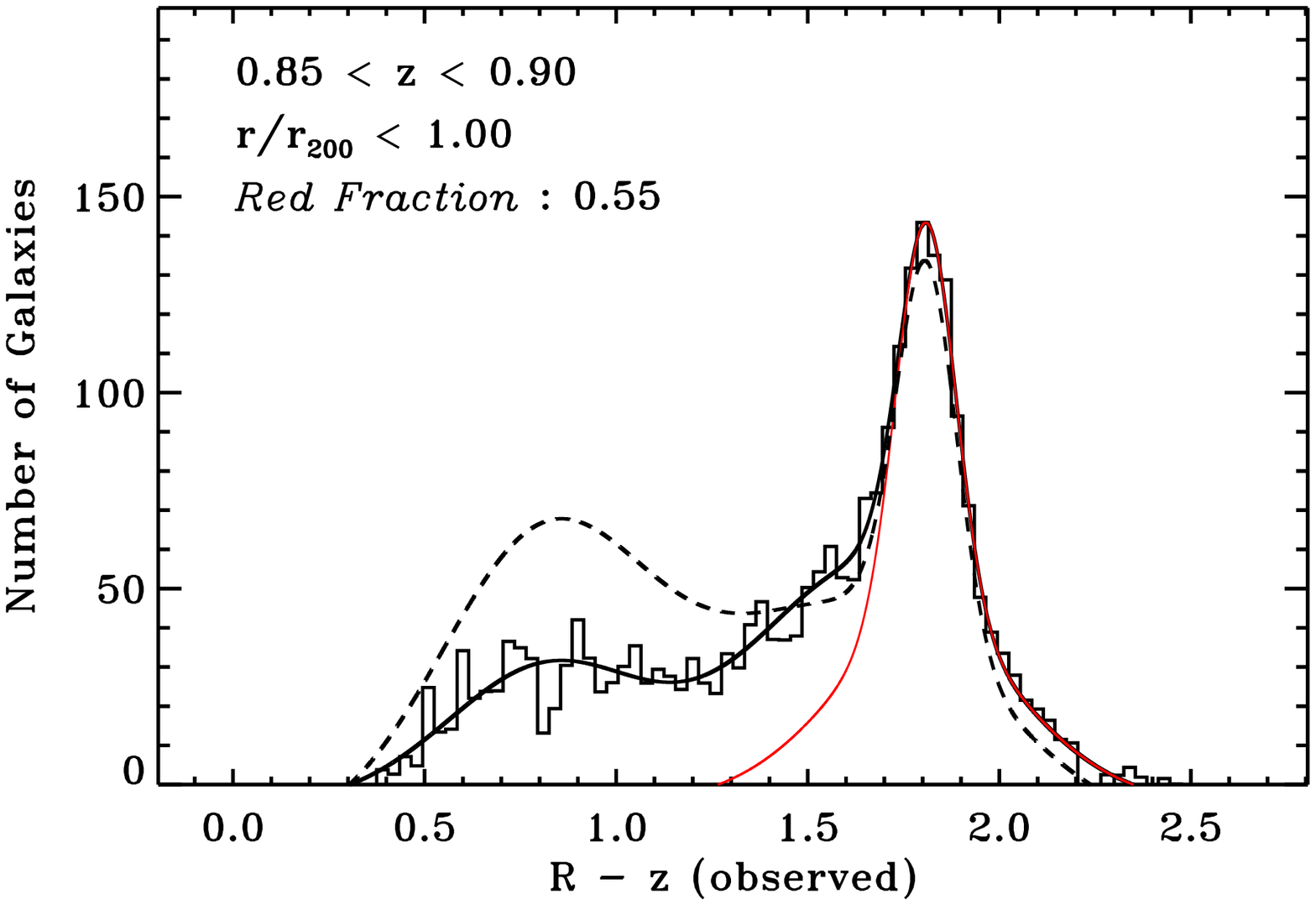}
\caption{The top panels are the background subtracted color-magnitude
distributions. The bottom panels are the color distribution after $k+e$ correction.
The histogram and the solid line shows the distribution after applying
a differential $k$-correction for non-red sequence galaxies (see section \ref{sec:diffk}) 
while the dashed lines show the distribution before such a correction was made.
The red curve is our model for the red distribution used to
compute the red fraction. The plots for are redshift slices:
$0.45 < \zph < 0.55$ (left), $0.65 < \zph < 0.75$ (middle) and $0.85 < \zph < 0.90$ (right). 
\label{fig:results}}
\end{figure*}

Here, we use an empirical approach that attempts to separate the red sequence galaxies from the
rest of the population, and divide the remaining galaxies amongst different fixed spectral types.
We first construct the $k+e$ corrected $R_c - z'$ color distribution by summing over the counts 
from all galaxies brighter than $M^\star +1.5$. The distribution for redshifts slices of
$0.45 < \zph < 0.55$, $0.65 < \zph < 0.75$ and $0.85 < \zph < 0.90$ 
are shown on the bottom panels of 
Figure \ref{fig:results}. We then fit the red side (split by the red sequence peak) of the 
distribution with a double Gaussian of a single mean, including a narrow peak and a broader
wing. Because the intrinsic spread of the red sequence is small 
\citep[$\lesssim 0.05$ {\tt mag};][]{Bow92,Blak03,Lop04,Coo05,Mei06} 
and there should be on average no excess of galaxies with redder colors associated with these clusters, this 
half of the color profile should be characteristic of the color uncertainty in our observations 
and should be symmetric about the red sequence peak. This peak is prominent enough at all observed 
redshifts that red sequence galaxies are thus robustly isolated.
The red fraction, $f_R = 1 - f_B$, the complement of the traditionally measured blue fraction, $f_B$, 
is estimated by mirroring the double Gaussian about the mean model distribution, and normalizing by 
the total galaxy counts (red and blue, smoothed with a spline curve). 
We note here that our approach is in spirit similar to the original \cite{But84} studies 
where the color width of their red sequence galaxies was indeed $\approx 0.2$ {\tt mag}.
This approach is also similar to the one suggested by \cite{And05} of using the valley 
between the two distributions as a divider \citep[\eg][]{Str01} and more recent bi-gaussian 
fitting to the local field color distribution (at fixed luminosity) by \cite{Bald04}. Our 
approach has the added advantage of estimating a more robust red fraction in the absence of a 
well-defined valley (e.g., the first distribution shown in Figure \ref{fig:results}), and when 
there is a departure from bi-gaussianity. Furthermore, the observed red distribution seems to broaden 
at increasing redshift. Our approach automatically captures this broadening, presumably due to the 
larger photometric uncertainties of the individual sources at fainter apparent magnitudes. Both 
the valley division and the fixed width approach will not take this into account, creating a 
susceptibility to an \citeauthor{Edd13}-like bias \citep{Edd13} 
which might mimic the redshift evolutionary trend we are trying to detect.
\subsection{Cluster-to-Cluster Variation}\label{sec:CtoC}
\begin{figure}  
\vspace*{1.25cm}  
\begin{center}
\includegraphics[width=0.48\textwidth,height=0.45\textwidth]{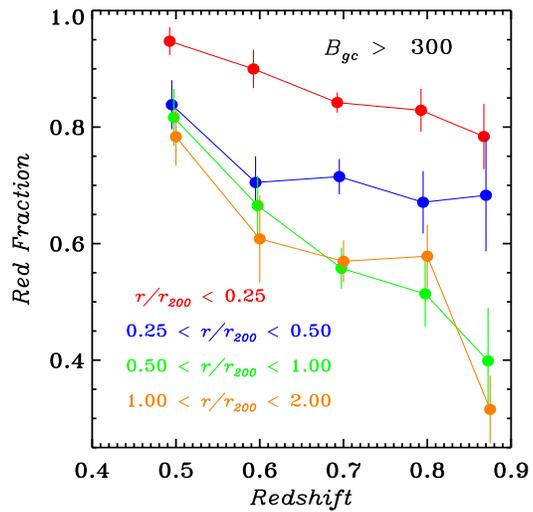}
\end{center}
\vspace*{0.25cm}  
\caption{The red fraction as a function of redshift for galaxies with
$r/\rvir < 0.25$ (core), $0.25 < r/\rvir < 0.5$, $0.5 < r/\rvir < 1.0$ and $1.0 < r/\rvir < 2.0$ (outskirts) and brighter than $M^\star + 1.5$.
\label{fig:res}}
\end{figure}

It has been noted from many previous studies of the Butcher-Oemler effect that even at fixed 
redshift, there is considerable variation in blue fraction. Some of this variation 
is due to the heterogeneous manner in which the clusters were selected, while others are intrinsic
in the sense that even at a given epoch, clusters have various physical morphologies 
(\eg~Bautz-Morgan Class) and states of evolution. 
We try to capture this variation between clusters by performing a jackknife 
analysis on our stacked clusters at a given redshift slice. This is done by 
dividing the cluster sample further into 15 to 20 subsamples, depending on 
the number of clusters involved. Each subsample is then removed, and the same 
red fraction analysis is performed. The error-bars displayed in 
Figure \ref{fig:res} are obtained by this approach. We note here that this 
approach also captures other variations within our data set, like small residual differences in patch to patch relative 
photometry. 
\section{Results and Discussion}\label{sec:discussion}
\subsection{The Butcher-Oemler Effect}\label{sec:BO}
The cumulative red fraction as a function of redshift is shown in Figure \ref{fig:res}. 
The fraction of red galaxies decreases as a function of redshift for all scaled 
radii $r/\rvir$. Particularly relevant to the Butcher-Oemler phenomenon is the
gradual decline in red fraction at the cluster core ($0.25~\rvir$) from $\sim 95$\%
at $z \approx 0.5$ to $\sim 80$\% at $z \approx 0.9$. 
The red fraction value of our lowest redshift bin is high compared with
the original \cite{But84} study and many subsequent analyses
which showed red fractions of about $0.7$ at $z \sim 0.4$.  Part of this
discrepancy comes from the way in which the red fraction is measured;
using a fixed color cut in the bimodal distribution will tend to
increase the blue fraction if the observational scatter is significant.
Our values are more consistent with recent studies using X-ray-selected cluster 
samples at $z \sim 0.3$ \citep{Wak05} and $z \sim 0.6$ \citep{And04}.

At face value, our result is inconsistent with the K-band analysis of \cite{Dep03} which probed 
a similar redshift range, but was based on a more massive (on average) and heterogeneous sample of $33$ optical, X-ray and 
radio-selected clusters. They fitted their data with a constant $f_B \sim 0.1$ out to 
$z \sim 0.9$. They confirmed the Butcher-Oemler effect as observed in an
optically-defined sample, but noted little evolution when using a
K-band selected sample. They concluded that any evolutionary effect may
be due to a population of bursting dwarf galaxies, whose K-band
luminosities (hence the stellar mass of the galaxies) are too faint for 
inclusion in a consistent K-band galaxy selection.

Our use of $k+e$ corrected magnitude in the $z'$-bandpass is not a perfect proxy for 
galaxy stellar mass but is less affected by recent star formation and has a much smaller $k+e$ 
correction compared with other optical bandpasses. Hence, we expect a reasonable agreement 
with K-band results. 
However, we find that our analyses differ significantly in the division of red versus blue galaxies.
Like \cite{But84}, \citeauthor{Dep03} use a fixed $\Delta(B-V) = 0.2$ {\tt mag} but $k$-corrected to their 
observed bandpasses using an effective SED derived from Sa + Sc mixture to give an effective 
$\Delta$(color) at higher redshift.
Unlike \citeauthor{But84} whose color histograms have an observed red sequence width of 
$\approx 0.2$ {\tt mag}, the color histograms from \citeauthor{Dep03} display a narrower red sequence 
width compared with the effective $\Delta$(color) -- often as large as $0.5$ {\tt mag} -- used in their analysis. 
Their chosen $\Delta$(color) allows the inclusion of galaxies of later spectral type in the red fraction,
suppressing their blue fraction measurements relative to ours, 
which model the observational scatter.
On examination of the individual color histograms of their high redshift clusters, 
we find that our method would produce substantial blue counts beyond the red sequence, 
and hence we find no evidence of inconsistency between our results and theirs.

We further note that the \citeauthor{Dep03}'s results are derived
using a fixed aperture of relatively small size ($0.25 - 0.45 \hmpc$),
whereas our red fractions are measured using apertures of scaled
radii $r/\rvir$.  Observations made within a fixed metric radius tend to sample
only the redder inner cores for massive clusters, and at $z>0.6$,
their sample is likely biased toward the most massive systems for
which the $\rvir$ values are typically as large as $1.5$ to $2.0 \hmpc$.
As Figure 6 for the RCS sample shown, at radii smaller than $0.5 \rvir$,
there is only a moderate evolution in the red fraction between $z \sim 0.4$
and $0.9$.


Our definition of red fraction is sensitive to our assumption that the red sequence color spread 
is symmetric, and dominated by observational uncertainties.
This appears to be the case for most of our data. However, in our two 
lowest redshift slices $0.45 < \zph < 0.50$ and $0.50 < \zph < 0.55$, we find that the chosen filters are not 
ideal separators of the red and blue distributions. 
In these two bins, the blue galaxies scatter underneath the red sequence 
so that our reflection of the red wing of the distribution will tend to oversubtract by a small amount. 
This effect is more problematic at fainter magnitudes because of the larger fraction of blue galaxies. 
For this redshift bin we fit the red sequence separately in four different magnitude bins to better 
separate the two distributions. Even so, the red fraction measured in this bin may be slightly too large 
by at most $5$\%. The problem only occurs 
in the lowest redshift slice, since at higher redshifts, the blue cloud is more widely separated in color 
space from the red peak for these filters. Inclusion of data from bluer filters will allow a better 
separation of the red and blue galaxies at the lower redshifts. An analysis using four color 
data \citep{Hsi05} over a smaller area will be explored in a future paper. 

\subsection{Radial Dependence}
A population gradient is seen for all redshifts in that the red fraction is larger in the 
cluster cores at all redshifts.
In Figure \ref{fig:res} we plot the red fraction as a function of redshift for galaxies 
within $r/\rvir < 0.25$, $0.25 < r/\rvir < 0.50$, $0.50 < r/\rvir < 1.00$ and $1.00 < r/\rvir < 2.00$. 
This result is similar to many low-redshift studies of 
clusters \citep[e.g.,][]{And06,Got04,Ell01}. In their analysis, \cite{Got04} found a break 
at $r/\rvir \sim 0.3-0.8$ at their highest redshift bin ($0.2 < \zph < 0.3$). 
Our results indicate a similar change in the radial gradient of the red fraction at
$r/\rvir \gtrsim 0.50$.
We caution such interpretations in light of the somewhat 
uncertain mass estimates of the clusters. Both statistical and systematic uncertainties in 
the cluster masses may affect our measures of the evolution of the red fraction and its radial 
dependence. Mass estimates from $\bgc$ measurements have $\sim 60$\% uncertainties for 
individual systems, which translate to $30$\% uncertainties in $\rvir$. Population gradients will be smoothed on this 
scale, possibly flattening any radial gradients on this scale, but most likely not affecting 
the comparison of evolution between $0.25\,\rvir$ and $1.00\,\rvir$. 

The decline in red fraction with redshift is significantly steeper in the outskirts
($1.00 < r/\rvir < 2.00$) compared to the core ($r/\rvir < 0.25$). Our results qualitatively 
agree with earlier work by \cite{Ell01} and \cite{Kod01} who found a milder decrease 
in red fraction at the core compared to the outskirts/field using an X-ray-selected clusters sample 
with spectroscopic redshifts. A possible interpretation is that this effect comes from a 
relative decline in the infall rates of bluer galaxies into clusters, as new additions to the 
cluster should generally lie at larger radii. We note that while systematic errors in our 
estimates of cluster masses, galaxy luminosity functions and $k$-corrections may affect our 
calculation of red fractions, this differential effect with respect to radius will remain, suggesting 
that the population in the cores of clusters changes with time differently from the 
population in the outskirts.

Our result on the radial dependence of the red fraction requires an accurate determination of each cluster center.
A detailed description for determining RCS cluster centers was given in \cite{Gla05}.
In brief, the location of a cluster is first selected from the color-
position data-cube with a magnitude weighting towards galaxies
brighter than $M^\star$; the positioning is then refined by an interative
Gaussian kernel of $250 \hkpc$. To the extent that the RCS algorithm is
accurate, then the centering should be better than $< 250 \hkpc$. Note
that clusters in reality are not round objects and often there are
double clusters, or galaxy overdensities along filamentary structures. If a
cluster is indeed elliptical and has a well-defined center, we argue
here that we will get the right center to better than half our
searching scale (i.e. $125 \hkpc$). If the cluster or galaxy
overdensity is filamentary, then our assumption of self-
similar spherical clusters breaks down. Indeed our results at $r < 0.25 \rvir$ maybe affected 
by centering problems, but not for the larger
radii. If we are not systematically deviating in our cluster center
determination, then the jackknife procedure in our error calculation
would capture the diversity and uncertainty of our red fraction
determination.

Because cluster galaxy populations show such marked radial gradients, any systematic variations of 
the $\bgc$ -- $\rvir$ relation with redshift may also change our red fraction as a function redshift, 
with any systematic tendency to overestimate cluster masses leading to a trend towards a lower red 
fraction. However, a significant systematic is required: even if our estimates of $\rvir$ are off 
by a factor of $3$ at $z \sim 0.8$, compared to $z \sim 0.5$, we would still see an evolution with redshift. 
Currently, no such redshift dependence is found in the RCS mass estimates when the same cluster 
sample is analysed using a global self-calibration procedure \citep[e.g.,][]{Hu03,Maj04}
which reflects standard cosmological parameters \citep{Gla06}. Ongoing work is currently being done 
to calibrate the mass of a subset of high redshift clusters using X-ray \citep{Hic07}, 
multi-object spectroscopy (\citealt{Gil07s}; Barrientos et al., in prep), weak lensing analyses (Hoekstra et al., in prep.) 
and mid-IR (Ellingson et al., in prep.) observations to constrain any such biases. 

\subsection{The Blue Galaxy Distribution}\label{sec:blue_gal}
\begin{figure*}  
\includegraphics[width=0.32\textwidth,height=0.32\textwidth,angle=90]{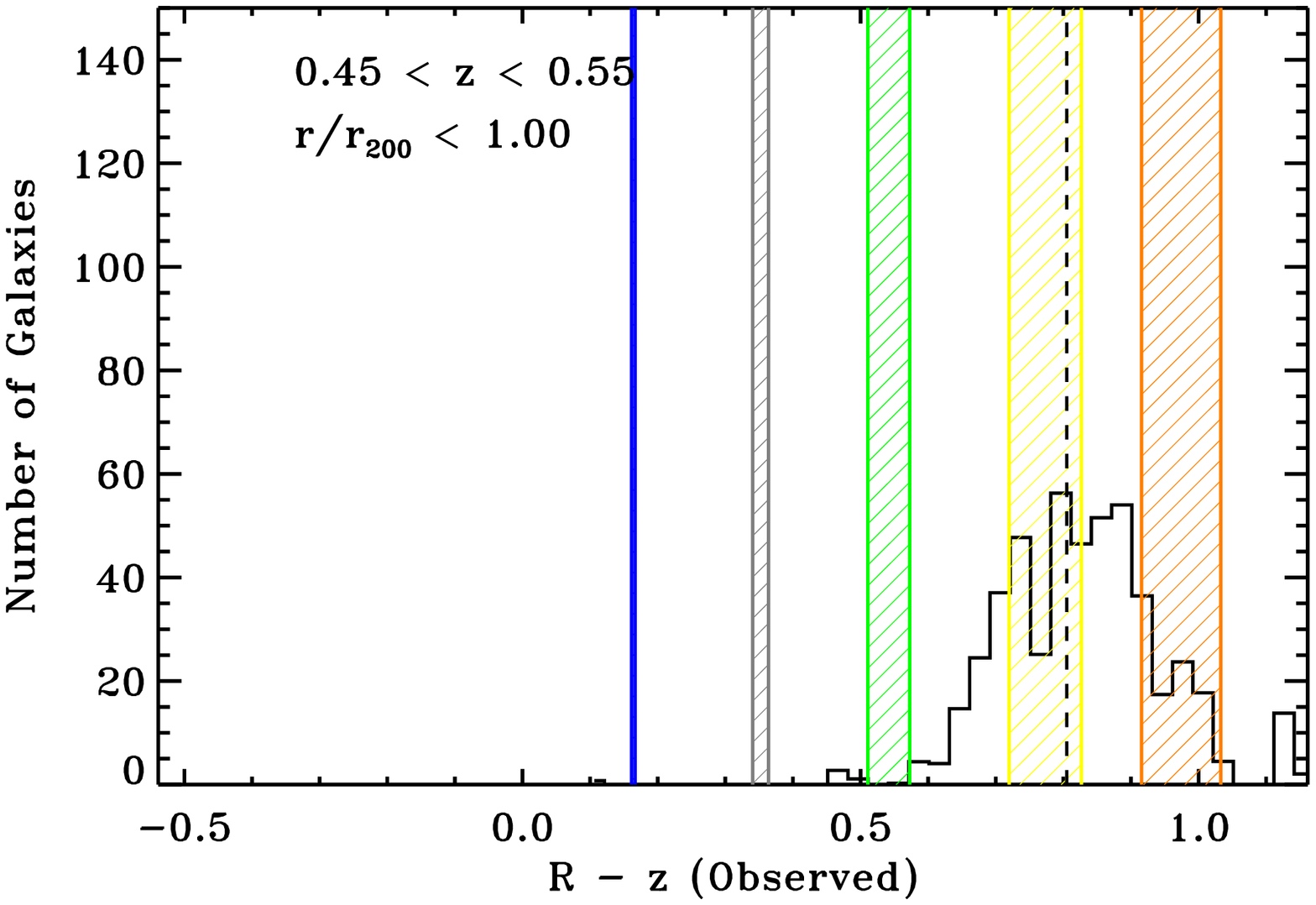}
\includegraphics[width=0.32\textwidth,height=0.32\textwidth,angle=90]{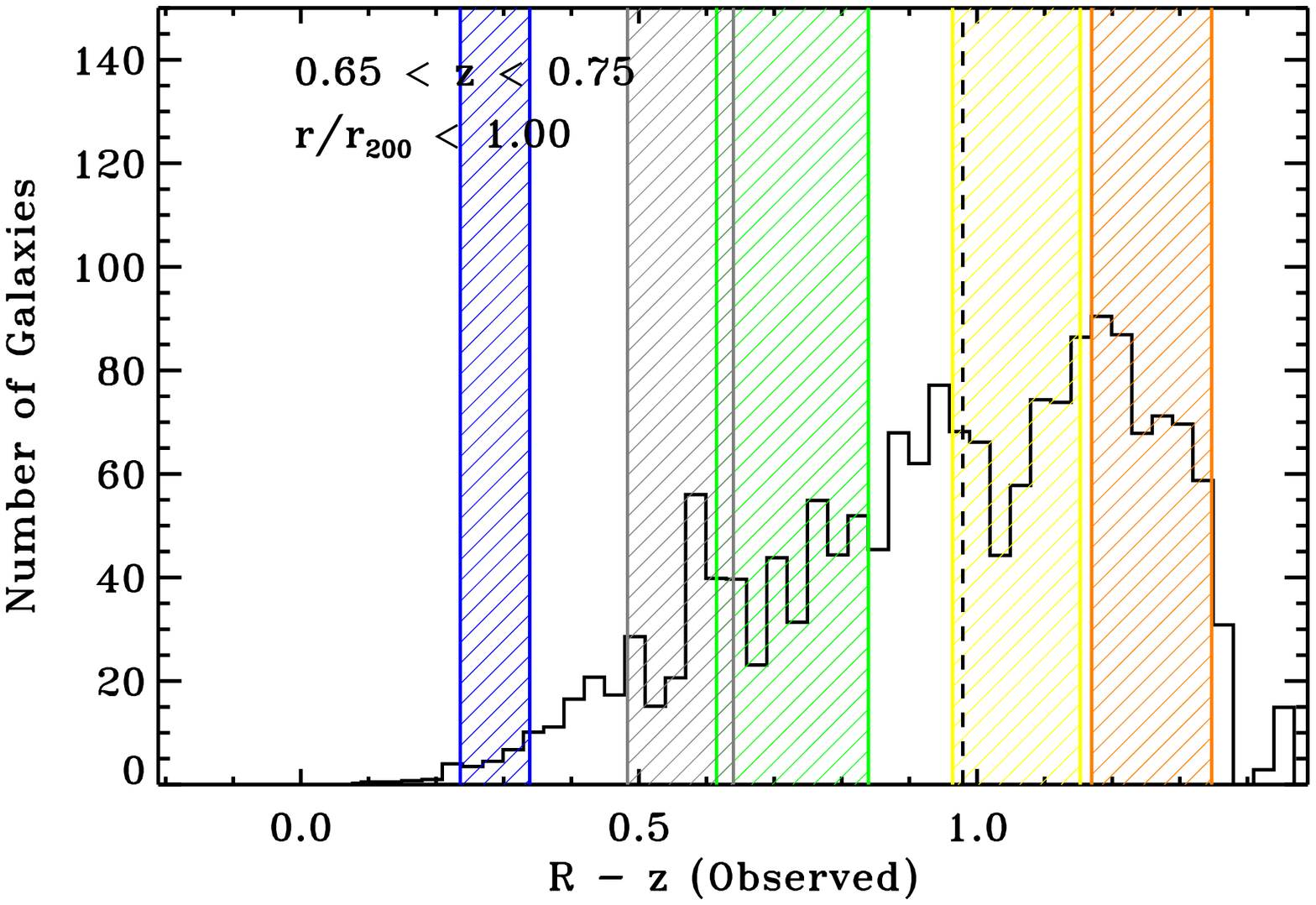}
\includegraphics[width=0.32\textwidth,height=0.32\textwidth,angle=90]{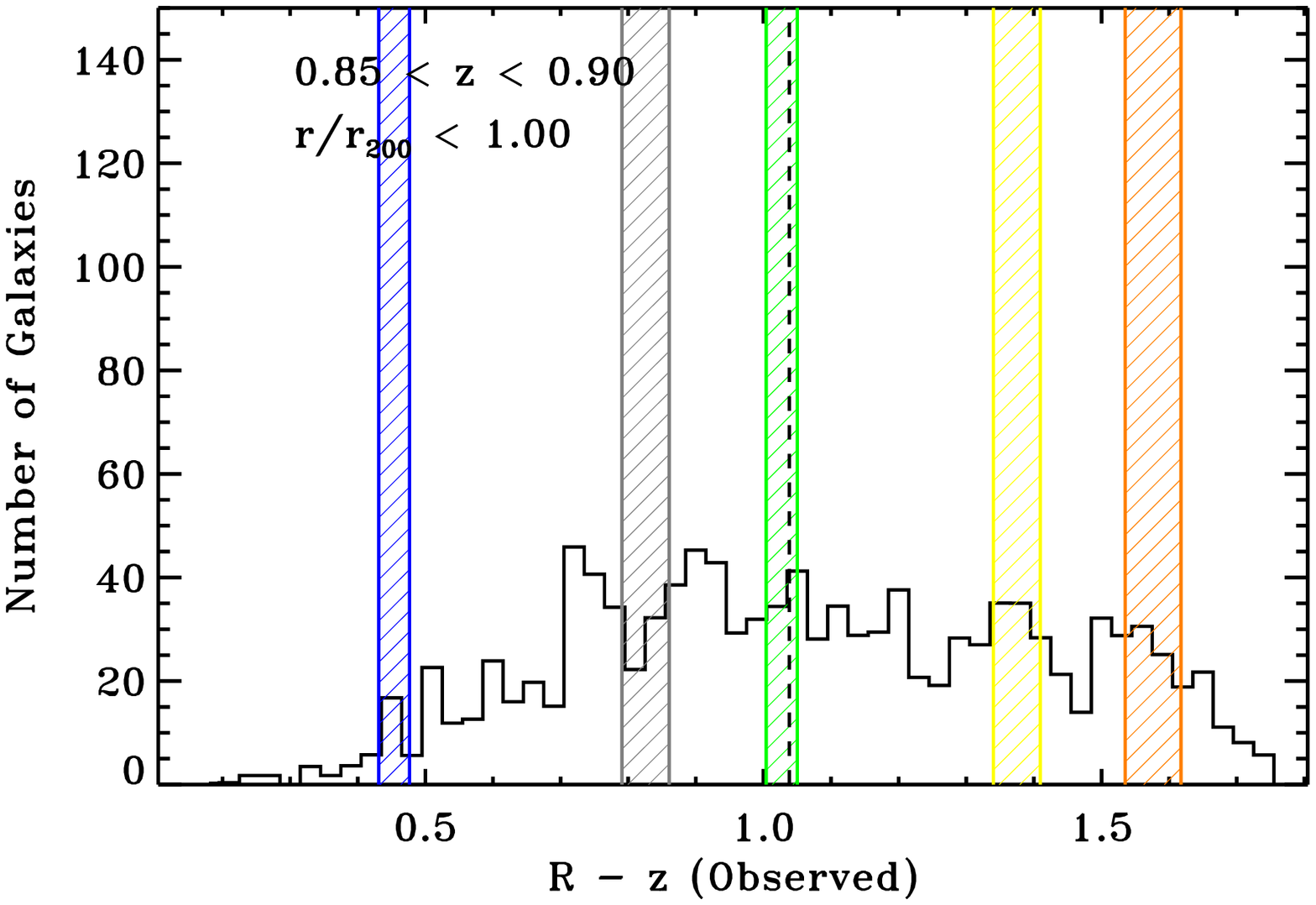}
\caption{The blue distributions of galaxies after correction for the red sequence (solid histograms). 
The shaded bands in each plot are the expected color for different spectral types, ranging from Sab (far right), Sbc,
Im, and starburst (SB1, SB2; far left), synthesized using a single representative SED for each type (see text). 
The width of the band reflects the finite redshift range in each subsample and the 
color variation as the filters shifts along the SED with redshift.
The plots are for redshift slices:
$0.45 < \zph < 0.55$ (left), $0.65 < \zph < 0.75$ (middle) and $0.85 < \zph < 0.90$ (right). 
The vertical dashed lines indicate the median colors of the distributions, also plotted in Figure 4.
\label{fig:bdist}}
\end{figure*}

Figure \ref{fig:bdist} shows the blue distribution of galaxies after removing the fitted red 
distribution. The shaded vertical bands are the expected color for galaxies of different (late) 
spectral type, ranging from Sab (far right) to starburst (SB; far left), synthesized using a single
representative SED for each type (cf. Figure \ref{fig:kcor}). The width of the band reflects the 
finite redshift range in each subsample and the color variation as the filters shift along the 
SED with redshift. The blue distribution, when compared with spectral type, appears to change slightly with 
redshift, becoming bluer at higher redshifts. The broadening towards the red tail at the highest redshift 
sample may be due to photometric uncertainty ($0.10-0.15$ {\tt mag}), but this cannot account for 
the longer tail to the blue. 
We find that the median galaxy spectral type for non-red-sequence galaxies changes by approximately 
half a spectral type between $z = 0.4$ and $z = 0.9$, (dashed line in Figure \ref{fig:bdist}, 
and the series of six blue squares in the middle panel of Figure \ref{fig:kcor}), and the 
fraction of blue galaxies with colors bluer than Im increases from $10$\% to $\approx 40$\%, 
between redshifts $0.5$ and $0.9$. Studies of color bimodality in field galaxies \citep{Str01,Bel04,Bla06}
suggest that field galaxies have colors consistent with slightly later types at these redshifts. 

The Butcher-Oemler effect has been attributed by some studies to a population of very blue 
starburst galaxies in clusters at high redshift \citep[e.g.,][]{Rak95,Dep03}. We investigate this 
possibilty by examining the redshift behavior of the fractions of blue galaxies with colors bluer 
then Im spectral type. The contribution from galaxies with spectral type similar to unobscured starbursts (we 
take the color mid-point between the the two models) is $< 5$\%, even for the highest 
redshift slice. This is an upper bound because we expect leakages from the more abundant redder 
galaxies to the blue tail due to photometric errors. Looking fainter into the luminosity function 
($> M^\star  + 1.5$) would increase the contribution from galaxies with later spectral types, but 
we show that the Butcher-Oemler effect as detected here is not driven by low-mass starbursts, 
and includes a component of brighter galaxies with colors consistent with normal coeval field galaxies.
Recently, works from \cite{Elb07} and \cite{Coo07} have found an 
increase in the number of bright blue, normal (i.e. not low mass or starburst) 
galaxies in high density environments at $z = 1$ relative to $z = 0$ (which they 
suggest may drive an inversion of the star-formation -- density relation).  Although 
our survey only probes rich clusters, our results suggest that the Butcher-Oemler effect is 
driven by normal galaxies, and are qualitatively consistent.
\subsection{Selection Effects from the Red Sequence Method}\label{sec:sel_eff_rs} 
By construction, our clusters are selected based on the red sequence
method using $z'$ and $R_c$ bandpasses, which will create a bias against 
finding clusters with very blue populations.  Our measurement of the 
Butcher-Oemler effect may thus be considered to be a lower limit. 
However, we do not expect that such incompleteness will create a large 
additional effect. Simulations of cluster detection efficiency using the RCS method
\citep{Gla02} show that the least massive clusters in our
sample (corresponding to $\bgc \approx 300 \bgcunits$) at $z \sim 0.9$  
are detected at high efficiency with red fractions as low as $0.55$. 
If the red fraction dips to only $0.2$, this efficiency drops to $50$\%. 
More massive clusters may be detected at $80$\% efficiency at a red fraction of $0.2$. 
Since we do not calculate red fractions for individual clusters here, we cannot
use this to create a quantitative correction. 
However, these
efficiencies are still significant even in the worst case scenarios,
and average values lie significantly far away from the red fractions where
we expect a selection bias to occur. 
Moreover, no other method for
finding high redshift clusters has identified a sample which contains a
large fraction of massive clusters with these very blue colors, and the overall
number of clusters found in the RCS survey appears to be consistent
with current cosmologies. Thus, we conclude that the problem of missing
very blue clusters of galaxies is not overwhelming, and the
Butcher-Oemler effect is not likely to be significantly stronger than
observed. Because of our red-sequence selection criterion, 
there is a possibility that our sample may have a slight relative bias
toward the redder systems at higher redshift.
If this is the case, then the observed 
Butcher-Oemler effect would be a slight underestimate of the true
evolution.
\section{Summary and Conclusions}\label{sec:summary}
We use $R_C$ and $z'$ observations of a sample of $\sim 1000$ clusters with masses greater 
than $\sim 10^{14} \hmsolar$ and $0.45 < \zph < 0.90$ from the Red-sequence Cluster Survey \citep{Gla05}  
to investigate the evolution of the color-magnitude relationship for cluster galaxies.  
We construct composite color-magnitude diagrams for a a number of redshift bins and ranges of 
cluster-centric radii, using cluster photometric redshifts and a statistical background subtraction 
technique.  We correct each galaxy color distribution for an empirically-determined estimate of the 
cluster luminosity function and $k$-corrections as a function of galaxy color, and investigate the 
bimodal distribution of excess galaxies in cluster fields. The large dataset and composite technique 
allows us to avoid many of the difficulties of previous investigations, where the luminosity and color 
evolution of different galaxy types and observational scatter blur the definitions of "red" and "blue" 
galaxies. Our analysis is informed by recent studies that the red sequence of galaxies is intrinsically 
narrow \citep[e.g.,][]{Bow92}, allowing us to fit an empirical description of the effects of 
observational scatter in our data, and separate galaxies in the red sequence from the rest of the 
cluster population.

We find a decrease in the fraction of red-sequence galaxies (increase in the blue fraction) in 
clusters across this redshift range, qualitatively consistent with the ”Butcher-Oemler effect.” 
Our measurements are difficult to compare quantitatively with previous measures of the blue fraction, 
and tend to yield somewhat higher red fractions because of our correction for scatter in the red 
sequence and explicit k-corrections for blue galaxies. Because of this, our results indicate a 
milder evolution than some earlier studies \citep[e.g.,][]{But84,Rak95}, but a stronger evolution
than that suggested by \cite{Dep03} using a small hetergeneous sample with K-band photometry. 
Our results are more consistent with recent studies of X-ray selected cluster samples 
\citep[e.g.,][]{Wak05}. 
We argue that while the red sequence technique may miss some fraction of very poor, blue clusters, 
this effect is minor, and the Butcher-Oemler effect is unlikely to be significantly stronger 
than seen here. We see consistent radial gradients in clusters, in that red fractions are higher 
in clusters cores than within larger radii at all redshifts studied. We also find a mild differential in 
the rate at which populations evolve -- more slowly within the inner cores of the clusters 
($r < 0.25\,\rvir$) than in the outer regions ($1.00 < r/\rvir < 2.00$), consistent with previous 
results at $0.3< z < 0.6$ \citep{Ell01,Kod01,And06}. This differential may indicate that 
cosmologically-driven infall of bluer galaxies into clusters may be partially responsible for 
the evolution in populations, as these galaxies will preferentially be found at larger radii. 
The decline in infall rates in low-density cosmologies then contributes to the decline in this 
population over time \citep{Ell03}.

The color distribution of non-red-sequence galaxies appears to be consistent with a 
population of normal spirals,  and shows a gradual evolution in the median color with redshift.  
There is only a minor contribution from galaxies with colors similar to unobscured starbursts. 
This places robust limits on interpretations of the Butcher-Oemler effect as being driven by 
starbursting dwarf galaxies that subsequently fade dramatically in clusters 
\citep[e.g.,][]{Rak95,Dep03}. However, the two-color data used here is insufficient to 
address quantitatively the star formation rates in cluster galaxies, or assess the possibility 
that obscured starburst and post-starburst galaxies play an important role in the evolution of 
cluster populations \citep{Pog99}. Additional studies of galaxy spectral energy distributions, 
morphologies and star formation rates are underway to trace the evolution of individual galaxies 
in clusters in this sample.

\acknowledgments
YSL and EE thank the National Science Foundation (NSF) grant AST-0206154 for support.
The RCS project is supported by grants from the Canada Research Chair Program, 
NSERC, and the University of Toronto to H.K.C.Y. 
LFB's research is partiatilly supported by FONDECYT under proyecto 1040423 and Centro de Astrofisica FONDAP.
YSL thanks Michael Strauss and John Stocke for discussions.
We thank Kris Blindert for sharing the results of her analysis of the mass-richness relation of 
intermediate redshift RCS clusters in advance of publication.



\onecolumn

\begin{table}
\begin{center}
\caption{\label{tab:sample}}
\begin{tabular}{r c c c c}
\hline \hline
redshift range&& Number of Clusters & \\ 
\hline
$0.45 < z < 0.55$  &&$171$ & \\
$0.55 < z < 0.65$  &&$203$ & \\
$0.65 < z < 0.75$  &&$229$ & \\
$0.75 < z < 0.85$  &&$263$ & \\
$0.85 < z < 0.90$  &&$118$ & \\
\hline
\end{tabular}
\end{center}
\end{table}

\end{document}